\documentclass[a4paper,12pt,fleqn,english]{article}

\usepackage[toc,page]{appendix}

\usepackage{a4}

\usepackage[latin1]{inputenc}
\usepackage[OT1]{fontenc}
\usepackage{babel,epsfig}
\usepackage{verbatim}
\usepackage{graphicx}
\usepackage{float}
\usepackage{epsf}
\usepackage{natbib}
\usepackage{amsmath}
\usepackage{amssymb}
\usepackage{amsfonts}
\usepackage{latexsym}
\usepackage{theorem}
\usepackage{dcolumn}

\usepackage{lineno}

\newcommand*\patchAmsMathEnvironmentForLineno[1]{%
  \expandafter\let\csname old#1\expandafter\endcsname\csname #1\endcsname
  \expandafter\let\csname oldend#1\expandafter\endcsname\csname end#1\endcsname
  \renewenvironment{#1}%
     {\linenomath\csname old#1\endcsname}%
     {\csname oldend#1\endcsname\endlinenomath}}%
\newcommand*\patchBothAmsMathEnvironmentsForLineno[1]{%
  \patchAmsMathEnvironmentForLineno{#1}%
  \patchAmsMathEnvironmentForLineno{#1*}}%
\AtBeginDocument{
\patchBothAmsMathEnvironmentsForLineno{equation}%
\patchBothAmsMathEnvironmentsForLineno{align}%
\patchBothAmsMathEnvironmentsForLineno{flalign}%
\patchBothAmsMathEnvironmentsForLineno{alignat}%
\patchBothAmsMathEnvironmentsForLineno{gather}%
\patchBothAmsMathEnvironmentsForLineno{multline}%
}

\graphicspath{{/nr/group/maler/nrdoc/}{./}}
\bibfont{\scshape}

\theoremstyle{plain}
\theoremheaderfont{\scshape}

\usepackage{array}
\makeatletter
\renewcommand\paragraph{\@startsection{paragraph}{4}{\z@}%
            {-2.5ex\@plus -1ex \@minus -.25ex}%
            {1.25ex \@plus .25ex}%
            {\normalfont\normalsize\bfseries}}
\makeatother
\begin{document}


\renewcommand{\textfraction}{0.0}

\title{A stage-structured Bayesian hierarchical model for salmon lice
  populations at individual salmon farms - Estimated from multiple
  farm data sets}
\author{Aldrin, M.$^{a*}$, Huseby, R.B.$^{a}$, Stien, A.$^{b}$, \\
  Gr{\o}ntvedt, R.N.$^{c}$,Viljugrein, H.$^{d}$, Jansen, P.A.$^{d}$\\
  \small $^a$  Norwegian Computing Center, P.O.Box 114 Blindern, NO-0314, Oslo, Norway\\
  \small $^b$ Norwegian Institute for Nature Research, P.O. Box 6606 Langnes, 9296 Troms{\o}, Norway\\
  \small $^c$ INAQ AS, P.O. Box 1223 Sluppen NO-7462, Trondheim, Norway\\
  \small $^d$ Norwegian Veterinary Institute, P.O. Box 750 Sentrum NO-0106, Oslo, Norway\\
  \small * Corresponding author: Tel: +47 22 85 26 58, fax: +47 22 69 76 60, \\
  \small Email address: magne.aldrin@nr.no}

\maketitle

\clearpage 

\textit{Abstract} Salmon farming has become a prosperous
international industry over the last decades. Along with growth in the
production farmed salmon, however, an increasing threat by pathogens
has emerged. Of special concern is the propagation and spread of the
salmon louse, \emph{Lepeophtheirus salmonis}. In order to gain insight
into this parasites population dynamics in large scale salmon farming
system, we present a fully mechanistic stage-structured population
model for the salmon louse, also allowing for complexities involved in
the hierarchical structure of full scale salmon farming. The model
estimates parameters controlling a wide range of processes, including
temperature dependent demographic rates, fish size and abundance
effects on louse transmission rates, effects sizes of various salmon
louse control measures, and distance based between farm transmission
rates. Model parameters were estimated from data including 32 salmon
farms, except the last production months for five farms which were
used to evaluate model predictions. We used a Bayesian estimation
approach, combining the prior distributions and the data likelihood
into a joint posterior distribution for all model parameters. The
model generated expected values that fitted the observed infection
levels of the chalimus, adult female and other mobile stages of salmon
lice, reasonably well. Predictions for the time periods not used for
fitting the model were also consistent with the observational data. We
argue that the present model for the population dynamics of the salmon
louse in aquaculture farm systems may contribute to resolve the
complexity of processes that drive that drive this host-parasite
relationship, and hence may improve strategies to control the parasite
in this production system.


\clearpage
\section{Introduction}  
\label{seq:introduction}

Salmon farming has become a large and economically
prosperous international industry over the last decades. Norway holds a
leading position as a producer of farmed salmonids with an annual
production of about 1.2 million tonnes, which is roughly half of the
worldwide production \citep{Anonymous2015}. Further growth in the
production of salmonids is in demand \citep{Anonymous2015} , but this
will come at the cost of increasing risks of pathogen propagation and
transmission. Large-scale host density dependence acting on pathogen
transmission has been demonstrated in salmon farming production
systems, both for macro parasites \citep{AldrinEtal2013b,
  JansenEtal2012, KristoffersenEtal2014} and viruses
\citep{AldrinEtal2011, AldrinEtal2010, KristoffersenEtal2009}. Of
special concern, is the propagation and spread of the salmon louse,
\emph{Lepeophtheirus salmonis}, and this parasite's potentially
harmful effect on wild salmon populations \citep{KrkosekEtal2007}.

Mathematical and statistical models are increasingly being used to
evaluate infection pathways and risk factors for pathogen propagation
and disease development, both in aquatic and terrestrial animal
farming \citep{AldrinEtal2013b, AldrinEtal2011, AldrinEtal2010,
  MurrayEtal2013, MurrayEtal2016, JonkersEtal2010, Diggle2006,
  Hoehle2009, KeelingEtal2001, ScheelEtal2007}. When such models are
able to describe the main patterns in the host-pathogen population
dynamics, including the spread within and between farms, they can be
used to predict future infection levels as well as simulate the
outcomes of disease mitigation scenarios, examples being interventions
to mitigate bovine tuberculosis in Great Britain
\citep{BrooksPollockEtal2014} and long term effects of infection
control measurements to mitigate salmonid alphavirus (SAV) incidences
causing pancreas disease (PD) outbreaks \citep{AldrinEtal2015}. The
Norwegian salmonid production system is exceptionally well suited for
developing models for salmon lice infection dynamics because of the
wealth of surveillance time-series that document both the spatial
locations and population sizes of host populations at risk of
infection, as well as salmon lice abundances in these host
populations. Coupling these host and parasite population data have
provided insights into e.g. how salmon lice spread between farms
depending on between-farm distances, and how transmission and parasite
abundances depend on local host biomasses \citep{JansenEtal2012,
  AldrinEtal2013b, KristoffersenEtal2014}. However, previous models
that describe both between and within farm parasite population
dynamics have for simplicity typically been autoregressive statistical
models focusing on single aggregated measures of parasite infection
levels \citep{ JansenEtal2012, AldrinEtal2013b, KristoffersenEtal2013,
  KristoffersenEtal2014}. Alternatively, models have been developed
for simulation purposes only \citep{GronerEtal2014} or focused on the
population dynamics on single farms over a limited time period
\citep{KrkosekEtal2010}. Most of these approaches have relied heavily
on estimates of demographic rates obtained in the laboratory
\citep{StienEtal2005, RevieEtal2005, GettinbyEtal2011, GronerEtal2013,
  RittenhouseEtal2016, GronerEtal2016}.

The aim of the present paper is to formulate a fully mechanistic
stage-structured population model for the salmon louse, that also
allows for the complexities involved in full scale salmon
farming. Furthermore, the model accounts for the hierarchical
structure of the data obtained from the production system where salmon
lice are counted on subsamples of fish, the fish being aggregated into
separate cages and the cages being aggregated to farm.  The model
estimates parameters controlling a wide range of processes, including
effects of temperature on demographic rates, fish size and abundance
effects on transmission rates, the different effect sizes, temporal
and stage specific effects of a wide range of salmon lice control
measures, and distance-based transmission rates between farms.  The
objectives for developing such a complex population model for the
salmon louse are: 1) To evaluate whether estimates of demographic
rates obtained in the laboratory seems applicable in full scale
production settings. 2) To evaluate the efficiency of different
control measures.  3) To explore importance of different sources of
infection (e.g. internal versus external sources).  4) To develop a
tool that can keep account of the salmon louse populations at the
production unit level in salmon farms, based on the successive
counting of salmon louse infestations. 5) To develop a tool for short
term predictions of salmon louse infection levels. 6) Finally, to
develop a sufficiently realistic model that can be used for
scenario-simulations exploring the effects of various parasite control
strategies. In this paper, we describe the model in detail and discuss
the results in relation to objectives 1-5.

\section{Materials and methods}
\label{sec:matmet}

\subsection{Modelling background}  
\label{subseq:mod.background}

Many authors have previously presented models for salmon louse
population dynamics. Most of these models are formulated on a
continuous time scale \citep{StienEtal2005, RevieEtal2005,
GettinbyEtal2011, GronerEtal2013, RittenhouseEtal2016,
KrkosekEtal2009} 
, whereas others are formulated at a discrete scale with a time step
one day \citep{GronerEtal2016}, which is also the case for our
model. However, more important when comparing with our model is that
the parameter values in previous models primarily are based on laboratory
data or on small scale experimental units in the marine
environment. When real production data has been used for estimation,
these have been aggregated. For instance, \cite{StienEtal2005} used
only laboratory data published in previous papers, whereas
\cite{GronerEtal2013} and \cite{GronerEtal2016} used values from
\cite{StienEtal2005} for some parameters and values from several other
previous papers for other parameters.  Furthermore,
\cite{RevieEtal2005} and \cite{GettinbyEtal2011} used laboratory data
from previous studies to estimate the majority of the parameters and
real production data to estimate the remaining parameters, but the
production data were aggregated over farms and to a monthly time
scale. One exception is the study by \cite{KrkosekEtal2009}, who
estimated their model by experimental data from small scale marine
cages, but their model covers only the parasitic stages of the salmon
louse.

The present estimating approach is fundamentally different. The
parameters are estimated by fitting the model to every lice count
collected through the whole production period on each cage at each of
32 farms. However, to ensure that the final parameter values are
within biological plausible ranges, we use laboratory data, mostly
based on results summarised in \cite{StienEtal2005}, to specify
informative prior distributions for many of the parameters.  The
priors are then updated to posterior distributions by the full scale
farm data using Bayesian methods. Since the model is estimated on real
data from many different farms under various conditions, it has to
simultaneously incorporate many features to handle activities or
events that affect lice abundance, including various types of
treatments, external infection from neighbouring farms and the
movement of fish (and then also lice) between cages at the same
farm. Our model is therefore more complex than the aforementioned
models.

For instance, our model takes into account and estimate the effect of
several different types of treatments, such as medical bath
treatments, in-feed treatments and the use of cleaner fish. Of the
aforementioned models, the model of \cite{GronerEtal2013} includes the
effect of cleaner fish, but the effect size was taken from previous
studies from the 1990-ies. Likewise, the models used in
\cite{RevieEtal2005}, \cite{GettinbyEtal2011} and
\cite{GronerEtal2013} include effects of medical treatments, but again
the values of the effects were based on previous studies. Our model is
designed to allow new interventions to be incorporated and their
effect on lice abundances to be estimated from real-time production
data.

Fish are always free of lice when they are stocked as smolt to
seawater cages, so the lice population at a farm is always initiated
by external infection. The models used in \cite{RevieEtal2005},
\cite{GettinbyEtal2011} and \cite{GronerEtal2013} include external
infection as a constant, estimated from aggregated data. In our model,
the external infection is farm-specific and time-varying, depending on
abundance of adult female lice at neighbouring farms.  Because we
intend to maximise the model fit to each lice count at each cage at
each farm, some of the parameters in our model are farm-specific or
even cage-specific and some are time-varying, whereas other parameters
are constant and common for all cages and farms.


\subsection{Salmon farming and salmon lice}
\label{subsec:salmonfarming}

Farm production of salmon comprises of a freshwater juvenile phase,
being followed by a marine grow out phase, the latter which is the focus of this
study. The production of salmon on a marine farm typically initiates
by stocking juvenile smolts to cages (or net-pens) either in spring or
in autumn.  Salmon are kept in the marine farms for about 1.5 years
after which they are slaughtered for food consumption. In Norway, only
fish of the same year class of age are kept on a given farm and we
term this a cohort throughout the present paper. After slaughtering,
it is mandatory to fallow the farm for a period of at least two months
before stocking a new cohort of salmon. Fish may occasionally be moved
from one marine farm location to an empty farm location, in which case
this farm will initially report fish weights larger than expected for
smolts recently stocked into the sea.

\subsection{Data}
\label{subsec:data}

\begin{figure}[htpb]
 \begin{center}
  \includegraphics[width=0.75\textwidth,angle=0]{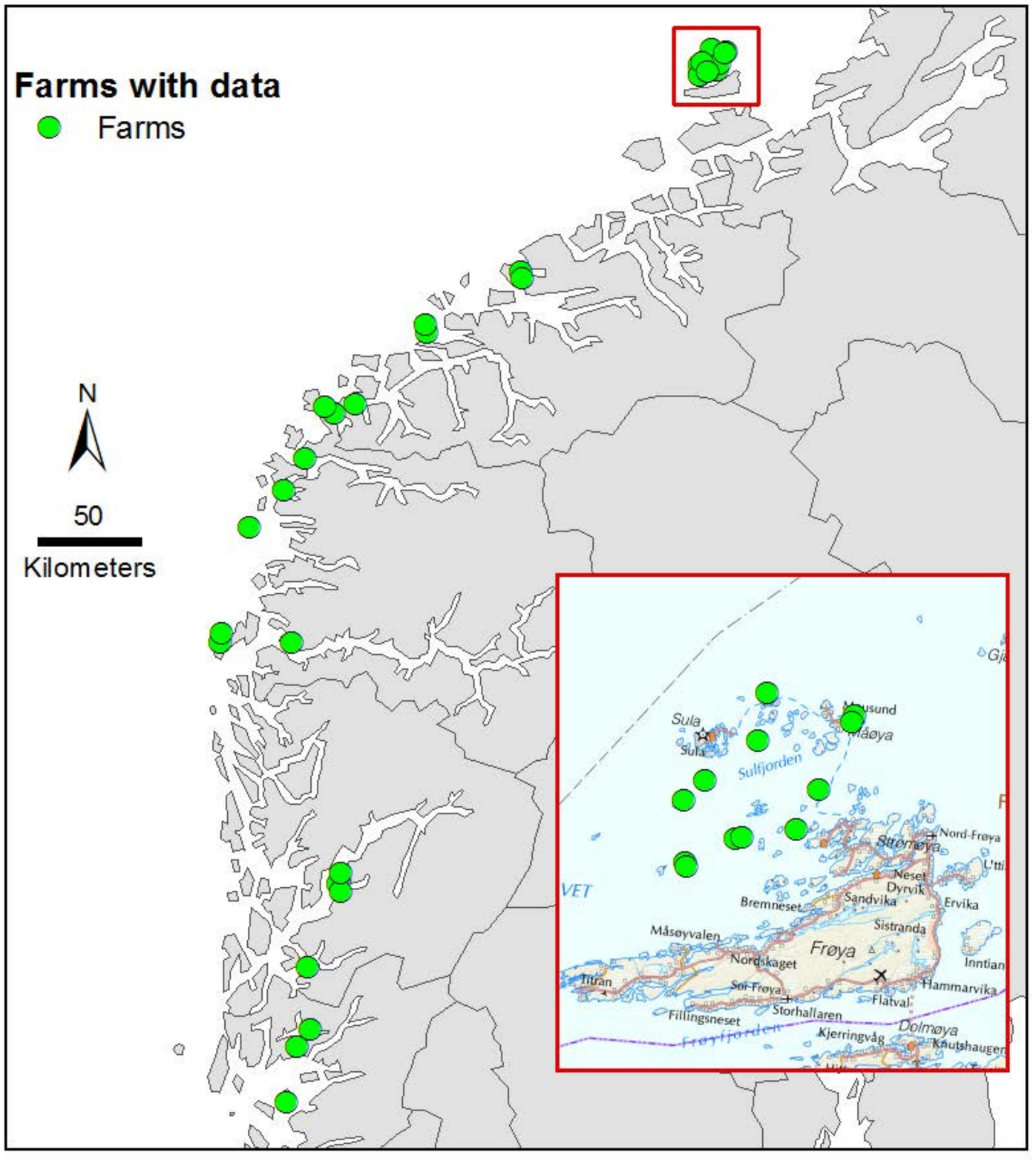}
  \caption{Geographical positions of the 32 salmon farms on the West
    coast of Norway (green circles). The highlighted area contains the
    12 farms in Nord-Fr{\o}ya.}
\label{fig:LocPos} 
\end{center}
\end{figure}

The main body of data in the present study consist of cage-level
data from 32 marine salmon farms in Norway, of which 12 farms are
located north of the island Fr{\o}ya in Mid-Norway (Figure
\ref{fig:LocPos}). For each farm, the data covers a full production
cycle for farmed salmon, from stocking as smolts to slaughtering as
adult Atlantic salmon (\emph{Salmo salar} L.), including fish
production data, lice counts, temperatures and louse control efforts.
Salmon were stocked between 2011 and 2013 and slaughtered about 1 1/2
year after stocking, between 2012 and 2014. The number of production
units (cages) per farm varied from 3 to 12, but were usually around 8
(mean 7.7). For 9 farms, the fish were moved between cages within
the farm during the production period.

Seawater temperatures were measured at 5 m depth at the farms.  The
average temperature was 9.1$^\circ$C, and 95 \% of the temperatures
were between 3.6 and 15.0$^\circ$C. Data on salinity were not available
in sufficient detail and have therefore not been used.

The production data consist of daily numbers and mean weights of
salmon per cage during the production period, information on movement
of salmon between cages within farms and information on antiparasitic
lice treatment using chemotherapeutic medicals (day of application and
type of medical).  Furthermore, the data contain information on
stocking of cleaner fish (day and number of cleaner fish stocked), but
with limited information on their mortality, and hence also for the
number of cleaner fish present at a given day.  We do not distinguish
between various species of cleaner fish.

The production cycles lasted on average 16.5 months per cage, with on
average 140 000 fish per cage, typically more in the beginning of a
production cycle and less towards the end. The average minimum and
maximum fish weights during a production cycle was 140 g and 5.7 kg,
respectively. 89 \% of the cages contained cleaner fish in
parts of the production period.

As a main rule, lice counts were performed on a sample of at least 10
fish every second week for each cage. The salmon lice were divided
into three categories according to developmental stages, \emph{i.e.}
i) chalimus (CH), ii) other mobiles (OM), which consist of pre-adults
and adult males, and iii) adult females (AF). There were on average 41
lice counts per cage, with averages (abundance) of 0.23 CH, 0.76 OM
and 0.18 AF per fish.

Six different types of antiparasitic medicals were used (Table
\ref{tab:medinfo}), and there were on average 4.6 events of medical
treatments per cage. We assume that the effects of deltamethrin and
cypermethrin are equal, since these are similar
compounds. Furthermore, when azamethiphos is used in combination with
deltamethrin or cypermethrin, we assume it has the same effect as
using deltamethrin or cypermethrin alone, since this combination is
used when reduced treatment effect is expected due to resistance
towards the medicals. The medicals emamectin benzoate and
diflubenzuron are given through the feed, typically over a period of
around two weeks. These treatments have a relatively low daily effect,
but effects last over a prolonged period. The other medicals are
applied as bath treatments over a duration of a few hours, with a
larger daily effect, but lasting over a shorter period. We assume it
is a time delay of $\Delta^{del}$ days (Table \ref{tab:medinfo}) after
application before the treatments give visible effects. Furthermore,
we assume that the duration of the effect, $\Delta^{dur}$, depends on
the seawater temperature according to
\begin{eqnarray}
\Delta^{dur} &=& \delta^{dur}/T_{t^0},
\end{eqnarray}
where $\delta^{dur}$ is a constant given in Table \ref{tab:medinfo}
and $T_{t^0}$ is the seawater temperature when the medical is applied. One
exception is when hydrogen peroxide was applied, for which
$\Delta^{dur}$ is temperature independent and given by
$\Delta^{dur}=\delta^{dur}$ (in days).

\begin{table}[h]
  \caption{Overview of types of medical treatments used. Codes: Y=yes, N=no, NA=missing information. The table content is based on  information in \cite{Nygaard2010} and \cite{OttesenEtal2012}. For hydrogen peroxide, the unit for $\delta^{dur}$ is days.}
  \footnotesize{
  \centering
\begin{tabular}{llrrccccc}
\hline
                        &          &                &Duration       &            &      &      &      &      \\
                        &          &Delay           &constant       &            &      &      &      &      \\
                        &Product   &$\Delta^{del}$   &$\delta^{dur}$ &Temperature &Effect&Effect&Effect&Effect\\
Medical                 &name      &(days)          &(days $\cdot ^{\circ}\mathrm{C}$)&dependency  &on CH &on PA &on A  &on egg\\ \hline
Deltamethrin             &Alphamax  &2               &84            &Y           &Y     &Y     &Y     &N     \\
Cypermethrin             &Betamax   &2               &84            &Y           &Y     &Y     &Y     &N     \\
Azamethiphos            &Salmosan  &1               &42            &Y           &N     &Y     &Y     &N     \\
Hydrogen peroxide (H$_2$O$_2$)&    &0               &7             &N           &N     &Y     &Y     &NA     \\
Emamectin benzoate      &Slice     &5               &210           &Y           &Y     &Y     &Y     &NA     \\
Diflubenzuron           &Releeze   &10              &126           &Y           &Y     &Y     &N     &N     \\ \hline
\end{tabular}
}
\label{tab:medinfo}\\
\end{table}

In addition to the detailed cage-level data on the 32 farms, we have more
aggregated data on all other Norwegian marine salmon farms. For a farm
$f^{\prime}$ at day $t$, we know the number of salmon, denoted by
$N_{tf^{\prime}}^{SAL}$. We also have an estimate
$\widehat{A}^{AF}_{tf^{\prime}}$ of the abundance of AF lice at the
farm, based on weekly lice counts on a sample of fish, and therefore also
an estimate of the total number of AF lice, given by
$\widehat{N}^{AF}_{tf^{\prime}}=\widehat{A}^{AF}_{tf^{\prime}}N_{tf^{\prime}}^{SAL}$.
Finally, we have the seaway distances between all farms, and we let
$d_{ff^{\prime}}$ denote the seaway distance between a farm $f$ and
another farm $f^{\prime}$.

Based on these quantities for all other farms, we have calculated an
external infection pressure index for each of the 32 farms, which can
be seen as a preliminary estimate of external infection pressure at
time $t$. This is a weighted sum of the estimated numbers of AF lice
at neighbouring farms, where the weights decrease by increasing seaway
distance to the farm in question. This index is denoted by
$N_{tf}^{AFExt}$ for farm $f$ at time $t$, and is given by
\begin{eqnarray}
\label{eq:PrelimInf}
N_{tf}^{AFExt} &=& \sum_{f^{\prime} \ne f} g(d_{ff^{\prime}}) \widehat{N}^{AF}_{tf^{\prime}} ,
\end{eqnarray}
where $g(\cdot)$ is a function decreasing by increasing distance given
by
\begin{eqnarray}
\label{eq:g}
g(d) &=& \exp(-0.618 d^{0.568}) .
\end{eqnarray}
This distance function is taken from \cite{AldrinEtal2013b}, and is
based on a data-driven model for lice abundance estimated from more
than eight years of data on all 1400 Norwegian salmon farms that were
active in the data period.

We have also calculated a corresponding weighted average of the
counted abundance of adult females at neighbouring farms, given by
\begin{eqnarray}
\label{eq:ab.ext}
A_{tf}^{AFExt} &=& \sum_{f^{\prime} \ne f} g(d_{ff^{\prime}}) \widehat{A}^{AF}_{tf^{\prime}} /\sum_{f^{\prime} \ne f} g(d_{ff^{\prime}}),
\end{eqnarray}

\begin{figure}[htpb]
 \begin{center}
  \includegraphics[width=1.2\textwidth,angle=0]{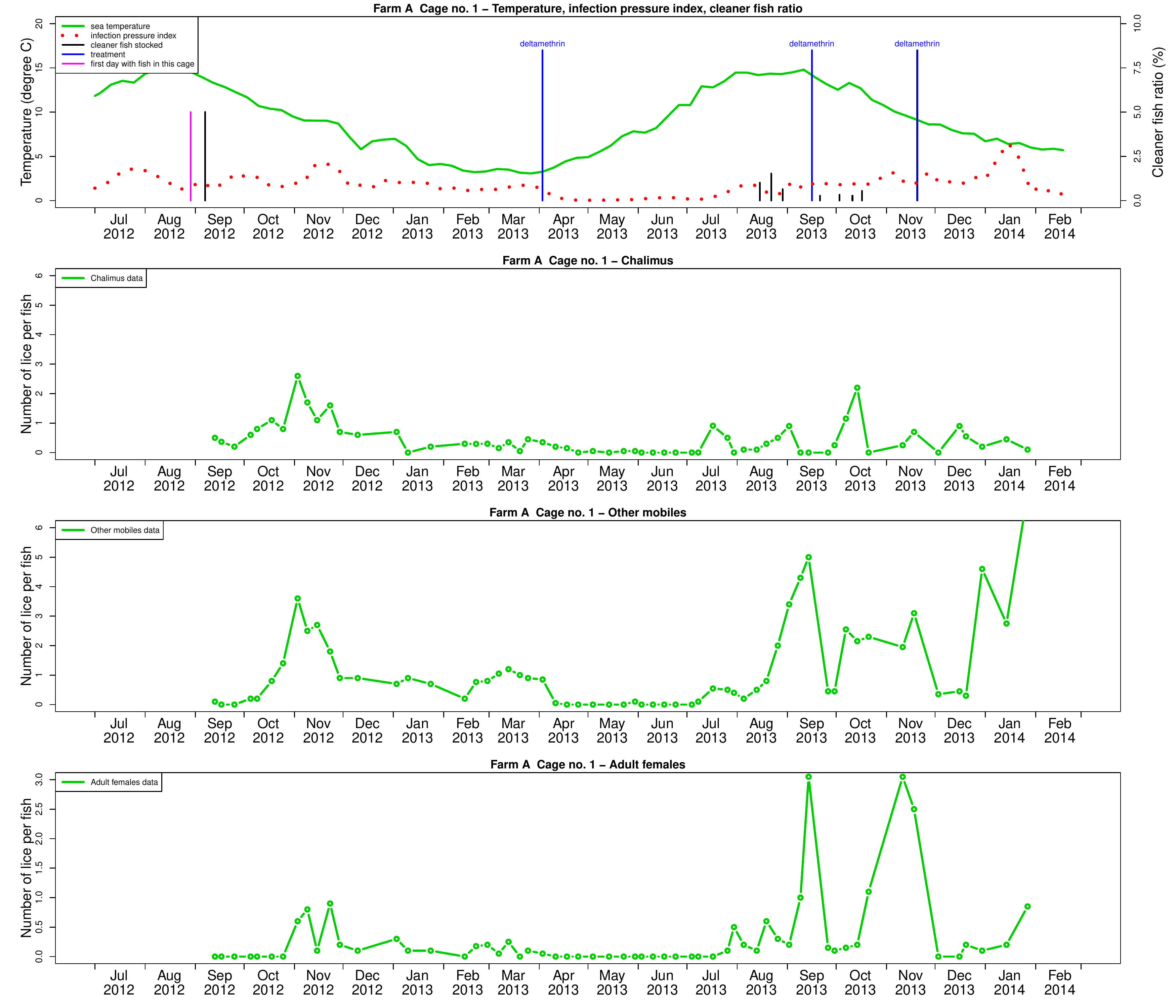}
  \caption{Counted lice abundance and other information for one cage
    at one farm. Upper panel: Seawater temperature (green line), external
    infection pressure index (red dotted curve), time of stocking
    (pink vertical line), treatments (blue vertical lines) and stocked
    cleaner fish ratio (black vertical lines). Three lower panels:
    Counts of chalimi, other mobiles or adult females shown as green
    circles connected by straight lines.}
\label{fig:FarmAdataonly} 
\end{center}
\end{figure}

Figure \ref{fig:FarmAdataonly} shows the most relevant data for one
cage at one farm. The upper panel shows time plots of the seawater
temperature (on the left y-axis) and the external infection pressure
index. In addition, the first stocking of salmon in this cage is
indicated by the vertical pink line and the various medical treatments
are shown as blue vertical lines. Finally, the stocking of cleaner
fish is also shown as vertical lines. The vertical extension of these
lines is proportional to the stocked cleaner fish ratio (on the right
y-axis), \emph{i.e.} the number of stocked cleaner fish divided by the number
of salmon.  The lower three panels show the counted abundance of lice
in the CH, OM and AF categories.

\clearpage
\subsection{Model framework}
\label{subsec:model}

\subsubsection{Some model overview and notation }
\label{subsubsec:overview}

Biologically, the life cycle of the salmon louse consists of eight
developmental stages \citep{HamreEtal2013}. These are aggregated into
the following five stages in our model: i) recruits (R, eggs and
nauplii larvae), ii) copepodids (CO, infective planktonic larvae),
iii) chalimi (CH, sessile lice on fish), iv) pre-adults (PA, mobile
lice on fish) and v) adults (A, also mobile lice on fish). The adults
are further divided into adult females (AF) and adult males (AM). When
an infective copepodid (stage CO) attaches to a fish host, it takes
approximately 24 hours before it moults into the CH stage
\citep{KrkosekEtal2009}. We ignore this short period and assume that a
copepodid enters the CH stage immediately upon attachment to a fish
host. The salmon lice count data from fish farms pool the stages PA
and AM together in one category (OM), whereas the stages CH and AF are
counted as separate categories.

The time resolution of the model is one day. The general idea is that
for stage-age $a=0$, the lice have developed into the given stage from
the previous stage, and that for $a>0$, the lice can develop into the
subsequent stage. We further assume that within a day, in the
following order;
\begin{itemize}
\item[0)] lice may be counted on a sample of fish,
\item[i)] lice may die due to natural mortality or treatment, 
\item[ii)] the surviving lice might develop to the next stage, and finally,
\item[iii)] fish, with sessile or mobile lice, can be moved to another
  cage or be slaughtered.
\end{itemize}

\begin{figure}[htpb]
 \begin{center}
  \includegraphics[width=1.2\textwidth,angle=0]{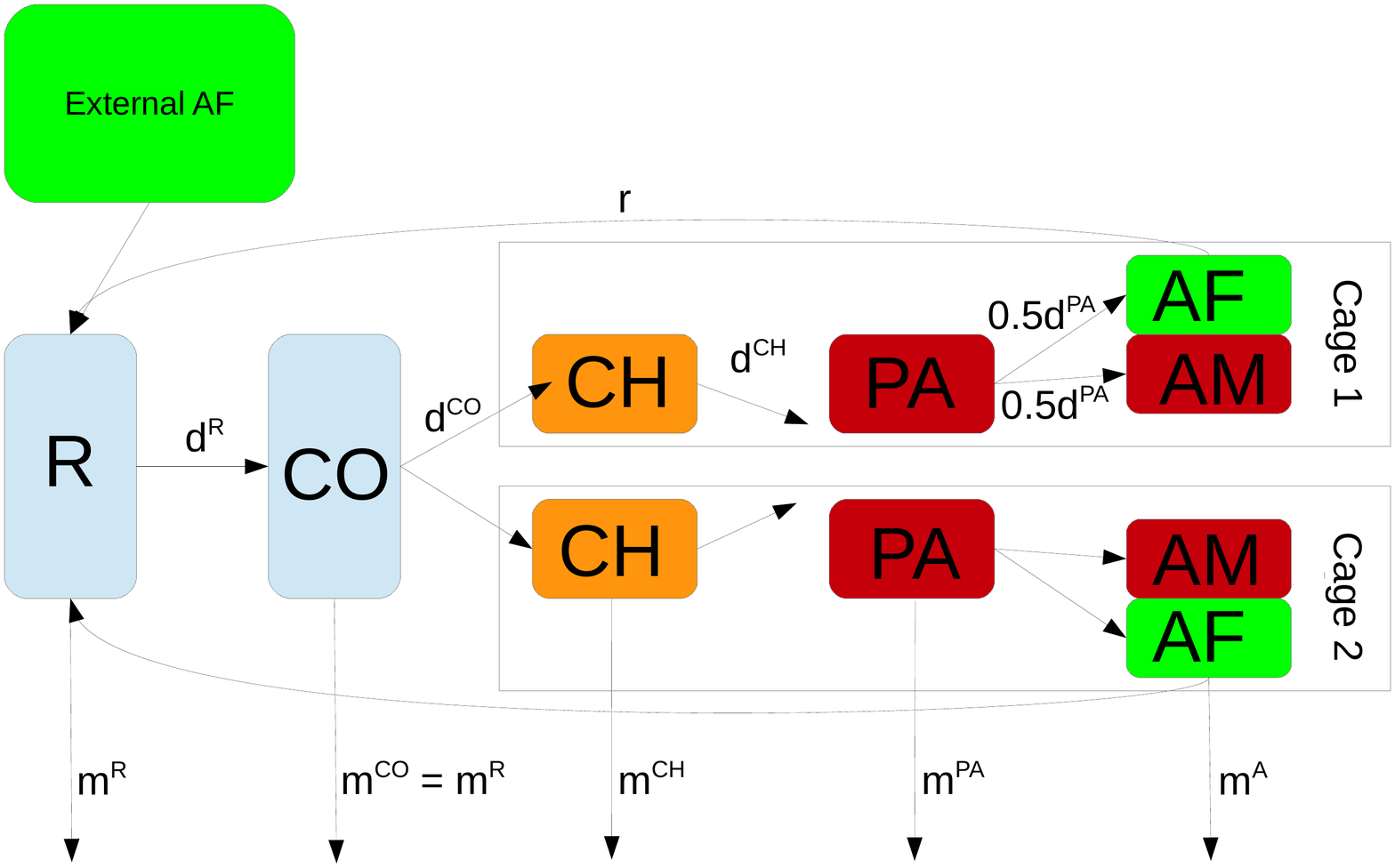}
  \caption{Overview of the population model for the salmon louse. Lice
    in the orange, red and green stages are counted, whereas lice in
    the blue stages are not counted. Lice are associated with a cage
    from the chalimus stage, here illustrated by a farm with two
    cages. The d-s, m-s and r-s symbolise development, mortality and
    recruitment, respectively.}
\label{fig:PopMod} 
\end{center}
\end{figure}

When fish are stocked to marine farms as smolt, they are free of lice,
and hence the initial lice transmission is caused by external
infections. Not until some lice at the farm have developed into
adults, the internal infection process can start. The population model
for a farm with two cages is illustrated by Figure
\ref{fig:PopMod}. In the R and CO stages, the lice are associated with
the farm, but not with any specific cage. From the CH stage and
onwards, however, the lice infect fish and are therefore associated
with specific cages. In the following subsections we describe the
various aspects of the population model and how the model is related
to lice count data. Table \ref{tab:notation} gives an overview of the
main notation we use.

The main elements of the model are inspired by the model in
\cite{StienEtal2005}, but with some extensions, e.g. to capture
management interventions on the farms. The model is validated rather
informally, by assessing whether parameter estimates are plausible and
by graphical evaluation of predictions ahead in time for five farms
where the last 3-11 months of data were not used in the estimation
process.

The model is estimated from the available data by a Bayesian
approach. We thus need prior distributions for the model
parameters. Many of these priors are informative, based on results
from laboratory experiments, \emph{e.g} those reported in
\cite{StienEtal2005}. For some priors, however, we use more vague
settings (see Section 2 in the Supplementary material for details).

\begin{table}[h]
  \caption{Overview of the model notation. When relevant, quantities may be used
    with subscripts $f$, $t$, $a$ and $c$, and with superscripts $R$,
    $CO$, $CH$, $PA$, $AF$ or $AM$} \footnotesize{ \centering
\begin{tabular}{|l|l|}
\hline
$f$& index for farm \\
$t$& index for time (day) \\
$a$& index for stage-age \\
$c$& index for cage \\
\hline
$N^{R}$& total number of lice recruits \\
$N^{CO}$& total number of copepodids  \\
$N^{CH}$& total number of chalimus larvae on fish in a given cage   \\
$N^{PA}$& total number of pre-adult lice on fish in a given cage   \\
$N^{AF}$& total number of adult female lice on fish in a given cage  \\
$N^{AM}$& total number of adult male lice on fish in a given cage  \\
\hline
$s$& survival rate (proportion per day)\\
$m$& mortality rate ($m=1-s$) \\
$d$& development rate (proportion per day) \\
$r$& reproduction rate (numbers per day and per AF lice)\\
$N^{SAL}$& number of salmon \\
$W$& average weight of salmon\\
$M_{c^{\prime}c}^{SAL}$& number of salmon moved from cage $c^{\prime}$ to cage $c$ \\
$w_{c^{\prime}c}$& proportion of salmon moved, $M_{c^{\prime}c}^{SAL}/N_{c^{\prime}}^{SAL}$ \\
$N^{CLF}$& number of cleaner fish \\
$S^{CLF}$& number of cleaner fish stocked \\
$Y$& Number of lice counted on a sample of $n$ fish\\
\hline
$\lambda$& parameters related to mortality\\
$\delta$& parameters related to development\\
$\mu$& expected values\\
$\beta$, $\gamma$, $\kappa$, $\rho$& various parameters\\
$\sigma^2$& variances\\
$z$& autoregressive processes\\
$\phi$& autoregressive coefficients\\
\hline
\end{tabular}
}
\label{tab:notation}\\
\end{table}

\subsubsection{Population model}
\label{subsubsec:popmodel}

Below, we present the population model for lice at a given farm (but
for simplicity without the farm index $f$). The model consists of two
equations per stage.  The first equation handles lice entering a given
stage at stage-age 0, typically from the preceding stage. The
exception being the R stage, where the new recruits are the result of
reproduction from adult females within the same farm and from
neighbouring farms. The second per-stage equation handles lice ageing
into higher stage-ages without developing into another stage.

The submodels for survival, development and reproduction
rates are presented later in Sections \ref{subsubsec:surv},
\ref{subsubsec:development}, \ref{subsubsec:developmentCO} and
\ref{subsubsec:reprod}.

{\bf{Model for the recruitment stage}} 

At the R stage, lice are not associated with a specific cage, but
rather seen as a reservoir of recruits with the potential to infect a
fish host at the farm in question in the future.  The model is:
\begin{eqnarray}
\label{eq:Rstage0}
N_{t(a=0)}^{R} &=& e_t^{Ext} N_{t-1}^{AFExt}r^{Ext}_{t-1} + 
\sum_c \sum_{a'} [N_{(t-1)a'c}^{AF} s_{(t-1)a'c}^{AF} r_{(t-1)a'c}] ,\\
\label{eq:Rstagea}
N_{t(a>0)}^{R} &=& N_{(t-1)(a-1)}^{R} s_{(t-1)(a-1)}^{R}  [1-d_{(t-1)(a-1)}^{R}] .
\end{eqnarray}

The first term in \eqref{eq:Rstage0} represents recruitment (into
stage-age 0) from neighbouring farms, also called external
recruitment. Here, $N_{t-1}^{AFExt}$ is a weighted sum of adult
females at neighbouring farms at time $t-1$ (see Section
\ref{subsec:data}). Furthermore, we assume that these reproduce with a
rate $r^{Ext}_{t-1}$ (see Section~\ref{subsubsec:external.recruitment}
). Then, $N_{t-1}^{AFExt}r^{Ext}_{(t-1)}$ can be interpreted as a
preliminary estimate of the number of external recruits reaching the
farm. However, this accounts for seaway distances to neighbouring
farms, but not for the sea currents in the area that may be more or
less favourable for a given farm, and which also may vary over
time. Therefore we have introduced the modifying factor $e_t^{Ext}$,
which is a farm-dependent and time varying modifying factor, see
Section \ref{subsubsec:omega.er} for an exact definition.

The second term in \eqref{eq:Rstage0} represents recruitment (into
stage-age 0) from adult female lice at the same farm, also called
internal recruitment. The product $N_{(t-1)a'c}^{AF}
s_{(t-1)a'c}^{AF}$ is the number of adult females at stage-age $a'$ in
cage $c$ that survives at time $t-1$ and they reproduce with a rate
$r_{(t-1)a'c}$ (see Section \ref{subsubsec:reprod}). The new recruits
are summed over all possible stage-ages of the adult females and over
all cages.

Eq. \eqref{eq:Rstagea} keeps track of the number of recruits of
stage-age $a>0$ that i) survives from the previous time point with
survival rate $s_{(t-1)(a-1)}^{R}$ (see Section \ref{subsubsec:surv})
and ii) do not develop into the infective CO stage, where
$d_{(t-1)(a-1)}^{R}$ is the development rate, i.e. the proportion of
recruits that develop into the CO stage (see Section
\ref{subsubsec:development}).

The equations for the next stages use similar notation for survival
rates, development rates and numbers of lice.


{\bf{Model for the copepodid stage}}
\begin{eqnarray}
\label{eq:COstage0}
  N_{t(a=0)}^{CO} &=& \sum_{a'} N_{(t-1)a'}^{R} s_{(t-1)a'}^{R} d_{(t-1)a'}^{R}, \\
\label{eq:COstagea}
  N_{t(a>0)}^{CO} &=& N_{(t-1)(a-1)}^{CO} s_{(t-1)(a-1)}^{CO}[1-\sum_{c}d_{(t-1)(a-1)c}^{CO}] .
\end{eqnarray}
Here, the development rate $d_{(t-1)(a-1)c}^{CO}$ (see Section
\ref{subsubsec:developmentCO}) represents the infection rate, i.e. the
proportion of the available copepodids that during a day infect fish
in cage $c$ and thus enter the CH stage. The sum over cages,
$d_{(t-1)(a-1)}^{CO}=\sum_{c}d_{(t-1)(a-1)c}^{CO}$ is then the total
infection rate at the farm. This is modelled as independent of
stage-age.

{\bf{Model for the chalimus stage}}
\begin{eqnarray}
\label{eq:CHstage0}
  N_{t(a=0)c}^{CH} &=& \sum_{a'} N_{(t-1)a'}^{CO} s_{(t-1)a'}^{CO} d_{(t-1)a'c}^{CO} ,\\
\label{eq:CHstagea}
  N_{t(a>0)c}^{CH} &=& \sum_{c'} N_{(t-1)(a-1)c'}^{CH} s_{(t-1)(a-1)c'}^{CH} [1-d_{(t-1)(a-1)}^{CH}] . 
\end{eqnarray}

From the CH stage on, the lice are attached to a fish, and therefore
associated with a specific cage.  We assume that the attached lice
follow the fish if the fish are moved to another cage or if the fish
are removed from the farm (including slaughtering and other fish
mortality). To handle this, the equations given here for the CH, PA
and AF stages are extended slightly (see Section 1.2 in the
Supplementary material).

{\bf{Model for the pre-adult stage}}
\begin{eqnarray}
\label{eq:PAstage0}
N_{t(a=0)c}^{PA}&=&  \sum_{a'} \sum_{c'} N_{(t-1)a'c'}^{CH} s_{(t-1)a'c'}^{CH} d_{(t-1)a'}^{CH} ,\\
\label{eq:PAstagea}
N_{t(a>0)c}^{PA}&=& \sum_{c'} N_{(t-1)(a-1)c'}^{PA} s_{(t-1)(a-1)c'}^{PA} [1-d_{(t-1)(a-1)}^{PA}] .
\end{eqnarray}

{\bf{Model for the adult stages}} 

For the adult stage, we distinguish in principle between males and
females. However, we assume that males and females have the same
survival and development rates and therefore each constitute 50 \% of
the adults. The main reason for this is that we do not have data on
the number of adult males on the fish, and therefore do not have the
information necessary for separate estimation of adult male
demographic rates.. The equations for adult females are then
\begin{eqnarray}
\label{eq:AFstage0}
N_{t(a=0)c}^{AF}&=&  0.5 \sum_{a'} N_{(t-1)a'c'}^{PA} s_{(t-1)a'c'}^{PA} d_{(t-1)a'}^{PA} ,\\
\label{eq:AFstagea}
N_{t(a>0)c}^{AF} &=& \sum_{c'} N_{(t-1)(a-1)c'}^{AF} s_{(t-1)(a-1)c'}^{AF} ,
\end{eqnarray}

while the number of adult males is equal to the number of adult females: 
\begin{eqnarray}
\label{eq:AFstage}
N_{tac}^{AM} &=&  N_{tac}^{AF} .
\end{eqnarray}

\subsubsection{Survival rates}
\label{subsubsec:surv}

We assume that the survival rates  may be farm-specific for some
stages, and therefore use the index $f$ when convenient, but we
sometimes drop the superscript that indicates the stage name.  We
assume that the total survival rate is the product of three terms;
\begin{eqnarray}
  s_{tfac} &=&  s_{tfac}^{nat} \cdot  s_{tfac}^{clf} \cdot s_{tfac}^{cht}= (1-m_{tfac}^{nat}) \cdot  (1-m_{tfac}^{clf}) \cdot (1-m_{tfac}^{cht}), 
\end{eqnarray}
where $s_{tfac}^{nat}$ is survival after natural mortality,
$s_{tfc}^{clf}$ is survival after additional mortality due to cleaner
fish predation (independent of stage-age) and $s_{tfc}^{cht}$ is
survival after additional mortality due to chemotherapeutic treatment
(independent of stage-age).  The $m$-s denote the corresponding
mortalities. The two latter terms are relevant only for the CH, PA and
A stages. All the three mortality (and survival) terms must lie
between 0 and 1, but they have different structures.

{\bf{Natural mortality}}

For the R and CO stages, we simply assume that the natural mortality is a
constant that is common for both stages, \emph{i.e.}
\begin{eqnarray}
  m_{tfac}^{Rnat}&=& m^{Rnat}=\lambda^{RCOnat} , \\
  m_{tfac}^{COnat}&=& m^{COnat}=\lambda^{RCOnat} 
\end{eqnarray}

For each of the CH, PA and A stages, we assume that the natural
mortalities are stochastic processes that can vary over time and
between farms, but are common for all cages within a farm and
independent of stage-age. This may account for factors that differ
between farms and change over time, for instance salinity, which is
not included in the model. Furthermore, including these mortalities as
farm-specific and time-varying terms improves the fit of the model to data. For
each farm and louse stage, the mortality is assumed to follow an
autoregressive model of order 1 (AR(1)) on the logit-scale as
\begin{eqnarray}
m_{tfac}^{nat} &=& m_{tf}^{nat} =\exp(z_{tf}^{nat}/(1+\exp(z_{tf}^{nat})), \\
\label{eq:mort.natCHPAAa}
(z_{tf}^{nat}-\lambda_{0}^{nat}) &=& \phi^{nat} \cdot (z_{(t-1)f}^{nat}-\lambda_{0}^{nat}) + \varepsilon^{nat}_{tf} ,\\
\label{eq:mort.natCHPAAb}
\mathrm{Var}(\varepsilon^{nat}_{tf}) &=& (\sigma^{nat})^2  .
\label{eq:mort.natCHPAAc}
\end{eqnarray}
Here, $z_{tf}^{nat} = \mbox{logit}(m_{tf}^{nat}) =
\log(m_{tf}^{nat}/(1-m_{tf}^{nat}))$. Furthermore, $\lambda_{0}^{nat}$
is the expected value on the logit-scale, $\phi^{nat}$ an
autoregressive coefficient and $\varepsilon^{nat}_{tf}$ a white noise
process with variance $(\sigma^{nat})^2$. These parameters have
separate values for each stage. In addition, the time-varying
mortalities $m_{tfac}^{nat}$ are restricted to lie within specified
intervals, which are (0.0006-0.02) for CH, (0.002-0.21) for PA and
(0.0003-0.70) for A. These limits are motivated from the various
studies summarised in \cite{StienEtal2005}, and are simply the most
extreme limits of the intervals given in their Table 4.

In addition, we assume $m$ = 1 from stage-age 80 for adults and from
stage-age 60 for the other stages. This is an approximation made to
save computer time.

{\bf{Mortality due to cleaner fish}}

We assume that cleaner fish feed on lice at the PA and A stages only
\citep{LeclercqEtal2014}, and that the corresponding mortality for
these two stages are equal.  Let
$x_{tfc}^{clf}=N_{tfc}^{CLF}/N_{tfc}^{SAL}$ be the ratio of the number
of cleaner fish to the number of salmon in cage $c$, at farm $f$ and
time $t$. This ratio depends among others on the mortality of cleaner
fish, which has to be estimated, and the model for this is described
later in Section~\ref{subsubsec:cleaner.fish}. Lice mortality in the
PA and A stages due to cleaner fish is then given by
\begin{eqnarray}
m_{tfac}^{clf} &=& m_{tfc}^{clf} =1-\exp(-\lambda^{clf}x_{tfc}^{clf}),
\end{eqnarray}
where the parameter $\lambda^{clf}$ is non-negative, such that the
mortality always is between 0 and 1.  One reason for assuming such a
simple model for the effect of cleaner fish (as opposed to the effect
of medical treatments discussed below) is that we also must estimate
the cleaner fish ratio which is multiplied by the cleaner fish effect.

{\bf{Mortality due to chemotherapeutic treatment}}

We assume that the chemotherapeutic treatments introduce extra
mortality of lice in some or all the stages CH, PA and A, depending on
the type of treatment. However, for simplicity we assume that lice are
only affected as long as they stay in the stage they were at the time
of treatment. If they manage to develop to the next stage, they are
clear of the treatment effect. Let the set of subscripts $fcbi$ denote
the $i$-th application of a chemotherapeutic of type $b$ in cage $c$
at farm $f$. Assuming that this treatment was given at time
$t_{fcbi}^0$, we define an indicator variable $x_{tfacbi}^{cht}$ that
is 1 when the treatment is active (a period after the treatment is
given) for lice at stage-age $a$, i.e. when
\begin{eqnarray}
t\in [t_{fcbi}^0+\Delta_{b}^{del},t_{fcbi}^0+\Delta_{b}^{del}+\Delta_{fcbi}^{dur})-1] \mbox{ and } a \ge t-t_{fcbi}^0,
\end{eqnarray}
where the delay constant $\Delta_{b}^{del}$ and the duration
$\Delta_{fcbi}^{dur}$ are explained in Section 2.3.

The mortality due to chemotherapeutic treatment is given by
\begin{eqnarray}
m_{tfac}^{cht} &=& 1-\exp(\sum_b -u_{fcbi}^{cht} x_{tfacbi}^{cht}),
\end{eqnarray}
where $u_{fcbi}^{cht}$ is a regression coefficient expressing the
effect of the specific application of the treatment. These regression
coefficients vary systematically between treatment types, accounting
for varying efficiency of different types of treatments. In addition,
they vary randomly between different applications of the same
treatment type, which for instance may be due to a varying degree of
resistance in the lice populations. This is handled by the following
formulation
\begin{eqnarray}
u_{fcbi}^{cht} &=& \log(1+\exp(u_{fcbi}^{cht*})) ,\\
u_{fcbi}^{cht*} &\sim& \mathrm{N}(\lambda^{cht},(\sigma^{cht})^2) .
\end{eqnarray}
In general, the parameters $\lambda^{cht}$ and $\sigma^{cht}$ differ
between various types of treatment, but are set equal for some
treatment types. These parameters are equal for the stages for which a
treatment have effect, but the effect of a specific treatment $fcbi$,
represented by the random coefficient $u_{fcbi}^{cht}$, may vary
between stages (this is for simplicity omitted from the notation above).

\subsubsection{Development rates}
\label{subsubsec:development}

We consider here the development rate from one stage to the next, for
the stages R, CH and PA. In all of the population models for lice that
we mentioned in the introduction, the development rate is 0 until
some, perhaps temperature-specific, minimum stage-age, and afterwards
positive and constant. In our opinion, the concept of a strict and
absolute minimum development time can be questioned in a population
with millions of individuals, and the assumption of a constant
development thereafter may be unrealistic. We have chosen to consider
the development to the next stage as a time-to-event process, and
model it as a discretised version of a Weibull distribution. The
Weibull distribution is widely used in statistical models for
time-to-event or survival analysis \citep{AalenEtal2008}. It also gave
a better fit to our data than a comparable formulation that included the
minimum development time model mentioned above as a special case (data
not shown).

When the time to an event is continiuous and Weibull distributed, the
event rate (often called hazard in survival analysis) is
$(\delta^{sc})^{-\delta^{s}}\delta^{s}a^{\delta^{s}-1}$, where $a$ is
the stage-age or time, $\delta^{s}$ is a shape parameter and
$\delta^{sc}$ is a scale parameter (sometimes
$(\delta^{sc})^{-\delta^{s}}$ is termed the scale parameter). In our
case, it is convenient to re-parameterise this as a function of the
median time to event, $\delta^m$, and the shape parameter. The event
rate then becomes
$\log(2)(\delta^m)^{-\delta^{s}}\delta^{s}a^{\delta^{s}-1}$, since the
median in the Weibull distribution is
$\delta^{sc}(\log(2))^{(1/\delta^{s})}$.

We use a discretised version of this, \emph{i.e.} our development rate is the
probability to develop to the next stage within a day, and it must
therefore also be restricted to be at most one.  We assume that the
median time to develop may vary over time and between farms, and
introduce therefore the subscripts $tfa$ on it. The model for the
development rate is then
\begin{eqnarray}
d_{tfa} &=& \min(\log(2)(\delta_{tfa}^m)^{-\delta^{s}}\delta^{s}a^{\delta^{s}-1},1) \quad \textrm{for} \quad a=0, 1, \ldots \quad.
\label{eq:dev.rate}
\end{eqnarray}

We further assume that the median development time depends on the
temperature history as $\delta_{tfa}^m =
c/(\bar{T}_{tfa})^{\delta^p}$, where $\bar{T}_{tfa}$ is the average
temperature that lice at stage-age $a$ at farm $f$ have experienced,
\emph{i.e.} the average temperature from time $t-a$ to time $t$, $c$
is a constant and $\delta^p$ is another constant that performs a power
transformation of $\bar{T}_{tfa}$. To get a more clear interpretation
of the constant $c$, we parametrise it as a function of the median
development time at 10$^{\circ}$ C, denoted by $\delta^{m10}$. The
final model for the median development time then becomes
\begin{eqnarray}
\delta_{tfa}^m &=& (10^{\delta^p}\delta^{m10})/(\bar{T}_{tfa})^{\delta^p} = \delta^{m10} (10/\bar{T}_{tfa})^{\delta^p} .
\label{eq:median.function}
\end{eqnarray}

The development rate defined by Eqs. \eqref{eq:dev.rate} and
\eqref{eq:median.function} also depends on the stage in the way that
the parameters $\delta^{m10}$, $\delta^s$ and $\delta^p$ are
stage-specific. One motivation for introducing $\delta^{m10}$ as a
basic parameter is that we use results on development times around
10$^\circ$ C from other studies as prior information, to ensure that
our estimates lies within biological plausible ranges. For the R
stage, which consists of eggs and nauplii, this prior information is
given separate for eggs and nauplii. Therefore, for the R stage,
$\delta^{Rm10}$ is the sum of one parameter $\delta^{Em10}$ for eggs
and another quantity $\delta^{Nm10}$ for nauplii, \emph{i.e.}
\begin{eqnarray}
 \delta^{Rm10}&=\delta^{Em10}+\delta^{Nm10}&  ,
\label{eq:medianR}
\end{eqnarray}
and $\delta^{Em10}$ is also contained in the reproduction factor
introduced later in Section \ref{subsubsec:reprod}.


In the estimation, we restrict $\delta^s$ to be larger than 1, and the
development rate will then be 0 at stage-age 0 and then increase by
increasing stage-age. Furthermore, the larger $\delta^s$ is, the more
steep will the development rate increase from 0 to 1 around the
median. When $\delta^s>2$, the difference between the mean and the
median will be less than 7 \%. It should further be noted that the
parameter $\delta^{m10}$ is only approximately the median development
time, since we consider a time-discrete version of the Weibull
distribution.

Assuming a constant temperature $T$, \cite{StienEtal2005} modelled the
minimum development time as $c_1/(T+c_2)^{c_3}$, where $c_1$, $c_2$
and $c_3$ are constants. They further assumed that $c_3=2$ and
estimated $c_1$ and $c_2$. We use a similar formulation for the median
development time, but assume $c_2=0$ and estimate $c_1$ and $c_3$. In
practice, these two formulations are quite similar for the relevant
temperatures and for the estimated values of $c_3=\delta^p$ (between 0.4
and 1.3, see Table Table \ref{tab:estmodel.static}).

\subsubsection{Infection rate}
\label{subsubsec:developmentCO}

The infection rate is the proportion of the copepodids that infect
fish during a day and thus develop into the CH stage. It is farm- and
cage-dependent, but does not depend on stage-age $a$, except that we
assume that development may only happen for $a \ge 1$.  This is
modelled as
\begin{eqnarray}
\label{eq:devCO}
d_{tfc}^{CO} &=& \exp(\eta_{tfc}^{CO})/(1 + \sum_c \exp(\eta_{tfc}^{CO})),
\end{eqnarray}
where 
\begin{eqnarray}
\label{eq:etaCO}
\eta_{tfc}^{CO}&=& \delta_{0fc}^{CO} + \log(N^{SAL}_{tfc}) +\delta_1^{CO}(\log(W_{tfc})-0.55) .
\end{eqnarray}
Here $N^{SAL}_{tfc}$ and $W_{tfc}$ are the number (in millions) and
the average weight (in kg), respectively, of fish in cage $c$ at farm
$f$ and time $t$, and 0.55 is roughly the mean of the natural
logarithm of the weight of fish. With this formulation,
$d_{tf}^{CO}=\sum_c d_{tfc}^{CO}$ will be the proportion of copepodids
that infect fish in any cage during day $t$, and this will always be
between 0 and 1. Furthermore, when the proportions or rates are small,
the rate $d_{tfac}^{CO}$ for each cage will approximately be
proportional to the number of fish $N^{SAL}_{tfc}$ in the cage and to
$W_{tfc}^{\delta_1^{CO}}$.

The parameter $\delta_{0fc}^{CO}$ controls the magnitude of the
infection rate conditioned on the number and weight of fish within a
given cage. In our model $\delta_{0fc}^{CO}$ depends on cage and farm,
reflecting that some farms or cages may be more exposed to infection
than others due to for instance sea current conditions. This is
handled by the following hierarchical structure:
\begin{eqnarray}
\delta_{0fc}^{CO} &\sim& \mathrm{N}(\delta_{0f}^{CO},(\sigma^{COdf})^2), \\
\delta_{0f}^{CO} &\sim& \mathrm{N}(\delta_{0}^{CO},(\sigma^{COd})^2) ,
\end{eqnarray}
where $\delta_{0f}^{CO}$ is a farm-specific mean and $\delta_{0}^{CO}$
an overall mean. Furthermore, $\sigma^{COdf}$ reflects the variability
between cages at the same farm, whereas $\sigma^{COd}$ reflects the
variability between farms.

\subsubsection{Reproduction factor}
\label{subsubsec:reprod}

The recruitment model, Eq. \eqref{eq:Rstage0}, includes the internal
reproduction factor $r_{tac}$, which is modelled taking into account
the following factors: Female adults extrude pairs of egg
strings. They can extrude a new set of egg strings within 24 hour
after the previous set was hatched, but hatching can take several days
\citep{StienEtal2005}. The number of eggs per string may increase for
each consequtive extrusion, which we approximate with
stage-age. Finally, not all eggs are viable. In addition, we allow for
density dependence in recruitment as suggested by
\cite{StormoenEtal2013}, \cite{KrkosekEtal2012} and
\cite{GronerEtal2014}.

The reproduction factor $r_{tac}$ for internal recruitment at time
$t$, stage-age $a$ and cage $c$ is thus modelled as
\begin{eqnarray}
\label{eq:rep}
r_{tac} &=& \beta^r_0 \cdot (a+1)^{\beta^r_1} \cdot 1/(\delta_{t}^{Em} +1) \cdot (1-\exp(-\gamma^r \cdot A_{tc})).
\end{eqnarray}
The first term in Eq. (\ref{eq:rep}), $\beta^r_0$, represents the
number of viable eggs for the first extrusion.  The next term,
$(a+1)^{\beta^r_1}$ models how the number of viable eggs per extrusion
increases by stage-age.  The third term, $1/(\delta_{t}^{Em} +1)$,
represents the rate of pairs of egg strings produced per day, which is
the inverse of average time between each egg extrusions, which further
is approximately the median hatching time plus one day for developing
new egg strings. The median hatching time is given by
\begin{eqnarray}
\label{eq:hatching.time}
\delta_{t}^{Em} &=& \delta^{Em10} (10/T_{t})^{\delta^{Rp}} ,
\end{eqnarray}
where $T_{t}$ is the seawater temperature and $\delta^{Em10}$ and
$\delta^{Rp}$ are parameters defined in Section
\ref{subsubsec:development}.  Finally, the term $(1-\exp(-\gamma^r
\cdot A_{tc}))$ allows for density dependent recruitment. Here,
$A_{tc}=N^{AF}_{tc}/N^{SAL}_{tc}$ is the abundance of adult females in
cage $c$ at time $t$.  A very large value of $\gamma^r$ corresponds to
a model without density dependent recruitment.  Of the parameters
involved in Eq. \eqref{eq:rep}, we estimate $\delta^{m10E}$,
$\delta^{pR}$ and $\gamma^r$ and fix $\beta^r_0$ and $\beta^r_1$ to
172.5 and 0.2, respectively (see Section 2.5 in the Supplementary
material for a motivation of these values).

\subsubsection{Reproduction factor for external recruitment}
\label{subsubsec:external.recruitment}

The reproduction factor $r^{Ext}_t$ for external recruitment in
Eq. \eqref{eq:Rstage0} is similar to the internal one, but the female
lice abundance $A_{tc}$ in Eq. \eqref{eq:rep} is replaced by a
weighted average of the counted abundance at neighbouring farms,
$A_{t}^{AFExt}$ (Section \ref{subsec:data} and Section 1.1 in the
Supplementary material).  Furthermore, we assume that all these female
lice at neighbouring farms are at stage-age $a=10$. The assumed
stage-age of 10 is rather arbitrary, but the results are insensitive
to this choice.

\subsection{Modifying factor in the external recruitment}
\label{subsubsec:omega.er}

The modifying factor $e_{t}^{Ext}$ for external recruitment is
farm-specific, so we include the farm index $f$ as well.  At the
log-scale, it varies over time around a farm-specific level according
to the following AR(1) model:
\begin{eqnarray}
e_{tf}^{Ext} &=& \exp(z_{tf}^{Ext}) ,\\
(z_{tf}^{Ext} - \mu_{f}^{Ext}) &=& \phi^{Ext} \cdot (z_{(t-1)f}^{Ext} - \mu_{f}^{Ext}) + \varepsilon_{tf}^{Ext} ,\\
(\varepsilon_{tf}^{Ext}) &\sim& \mathrm{N}(0,(\sigma^{Extar})^2) ,\\
\mu_f^{Ext} &\sim& \mathrm{N}(\mu^{Ext},(\sigma^{Ext})^2) .
\end{eqnarray}
Here, $\mu_{f}^{Ext}$ is the farm-specific expected value on the
log-scale, $\phi^{Ext}$ the autoregressive coefficient and
$(\sigma^{Extar})^2$ the residual variance. Furthermore, $\mu^{Ext}$ is
the overall expected value and $(\sigma^{Ext})^2$ the between-farm
variance of $\mu_{f}^{Ext}$.

\subsubsection{Cleaner fish model}
\label{subsubsec:cleaner.fish}

Let $S^{clf}_{tc}$ denote the number of cleaner fish stocked and
$N^{clf}_{tc}$ the total number of cleaner fish in cage $c$ at time
$t$. $S^{clf}_{tc}$ is observed, whereas $N^{clf}_{tc}$ is unknown and
modelled as
\begin{eqnarray}
N^{clf}_{tc} &=& N^{clf}_{(t-1)c} (1-\kappa^{clf}) + S^{clf}_{tc} ,
\end{eqnarray}
where $\kappa^{clf}$ is the daily constant mortality rate of cleaner
fish, common for all farms.

\subsubsection{Data model}
\label{subsubsec:data.model}

In this subsection, we describe how the population model is related to
the lice count data.  Let $Y_{tc}^{CG}$ be the number of lice in count
group $CG$ found on $n_{tc}$ counted fish at time $t$ and cage $c$,
where the count groups are either chalimus ($CH$), adult females
($AF$) or other mobiles ($OM$, \emph{i.e}, pre-adults and adult
males). We assume that these follow a negative binomial distribution
with mean $\mu_{tc}^{CG}=\mathrm{E}(Y_{tc}^{CG})$ and a heterogeneity
or aggregation parameter $n_{tc}\rho^{CG}$, such that the variance of
$Y_{tc}^{CG}$ is
$\mu_{tc}^{CG}+(\mu_{tc}^{CG})^2/(n_{tc}\rho^{CG})$. Deleting the
superscript $CG$ and subscript $tc$ for a moment, the probability
distribution of $Y$ is
\begin{eqnarray}
\label{eq:NBdist}
P(Y=y) &=&\frac{\Gamma(y+n\rho)}{y!\Gamma(n\rho)}(\frac{n\rho}{n\rho+\mu})^{n\rho} (\frac{\mu}{n\rho+\mu})^y .
\end{eqnarray}
We get the total likelihood for each count group by multiplying over
all counts, cages and farms. We further assume independence between
count groups and get the total likelihood by multiplying the
contribution from each count group.

The expected numbers of the various $Y_{tc}^{CG}$'s are given from the
population model as
\begin{eqnarray}
\mathrm{E}(Y_{tc}^{CH}) &=& n_{tc} \cdot p^{CHcount}_{tc}  \cdot N_{tc}^{CH}/N^{SAL}_{tc} ,\\
\mathrm{E}(Y_{tc}^{AF}) &=& n_{tc} \cdot N_{tc}^{AF}/N^{SAL}_{tc} ,\\
\mathrm{E}(Y_{tc}^{OM}) &=& n_{tc} \cdot (N_{tc}^{PA} + N_{tc }^{AM})/N^{SAL}_{tc} ,
\end{eqnarray}
where the role of the factor $p^{CHcount}_{tc}$ is to adjust for
under-reporting of CH lice, since they are very small and difficult to
count, especially on large fish.  We assume that this factor is
farm-specific (for instance, the staff at some farms may be more
trained or motivated than staff at other farms), and we introduce from
now on the index $f$ for farm. Then, the model for $p^{CHcount}_{tfc}$
is
\begin{eqnarray}
p^{CHcount}_{tfc} &=& \exp(\eta_{tfc}^{CHcount})/(1+\exp(\eta_{tfc}^{CHcount})),
\end{eqnarray}
where 
\begin{eqnarray}
\eta_{tfc}^{CHcount} &=& \beta^{CHcount}_{0f} + \beta^{CHcount}_1 (W_{ftc}-0.1) ,
\end{eqnarray}
where $W_{ftc}$ as before is the mean weight of fish in cage $c$ at
farm $f$ at time $t$. The constant 0.1 is chosen to make it easier to
specify prior distributions for $\beta_{0f}^{CHcount}$ and
$\beta_1^{CHcount}$. Here, $\beta_1^{CHcount}$ is common for all
farms, but $\beta_{0f}^{CHcount}$ varies between farms according to
the following hierarchical model:
\begin{eqnarray}
\beta_{0f}^{CHcount} &\sim& \mathrm{N}(\beta_{0}^{CHcount},(\sigma^{CHcount})^2) .
\end{eqnarray}

The model was estimated from the data including 32 farms, except the
last months (3-11) of data for five of the farms that were used for
evaluating conditional predictions. We used a Bayesian estimation
approach, combining the prior distributions and the data likelihood
into a joint posterior distribution for all model parameters. This was
done by Markov Chain Monte Carlo (MCMC) simulations
\citep{GilksEtal1996}. First, several initial chains were run to
identify a rough range for plausible parameter values.  Then four
independent chains were started from slightly different starting
values within this range. The first 25000 iterations were used as
burn-in to establish convergence, and the posterior distributions were
calculated by combining 100 thinned samples from the last 6000
iterations from each of the chains. See Section 4 in the Supplementary
Material for more details on the MCMC algorithm.

\clearpage
\section{Results and discussion}
\label{sec:results}

\subsection{Fitted and predicted values}
\label{subsec:estmodel}

The model generated expected values that fitted the observed infection
levels of chalimus (CH), adult female (AF) and other mobile stages
(OM) well (Figure~\ref{fig:FarmAFittedPredicted}, and Section 3 in the
Supplementary material with results for seven other
farms). Predictions for the time periods not used for fitting the
model were also consistent with the data with respect to the timing of
population growth of adult female (AF) and other mobile stages (OM)
(Figure~\ref{fig:FarmAFittedPredicted}, and Figure 1-5 in Section 3 in
the Supplementary material). These results support the notion that
there is a substantial deterministic component in the transmission
pathways and population dynamics of salmon lice in fish
farms. 

However, for periods with elevated predicted population sizes,
abundances of salmon lice were sometimes over-estimated (e.g. AF
abundance in August 2013, Figure 3, Section 3 in Supplementary
material) and sometimes under-estimated (e.g. AF and OM in first part
of September 2013, Figure \ref{fig:FarmAFittedPredicted}). These large
deviations in some predictions are likely to reflect 1) that there are
predictor variables that have not been included in the present model
(e.g. salinity), 2) substantial uncertainty in some predictor
variables like the abundance of cleaner fish in the cages, and 3) that
stochasticity, in particular with respect to the infection process,
limit our ability to make precise predictions. Accordingly, also the
credible intervals for the predictions were wide when elevated
abundances of infection were predicted (e.g. Figure
\ref{fig:FarmAFittedPredicted}).

\begin{figure}[htpb]
 \begin{center}
  \includegraphics[width=1.2\textwidth,angle=0]{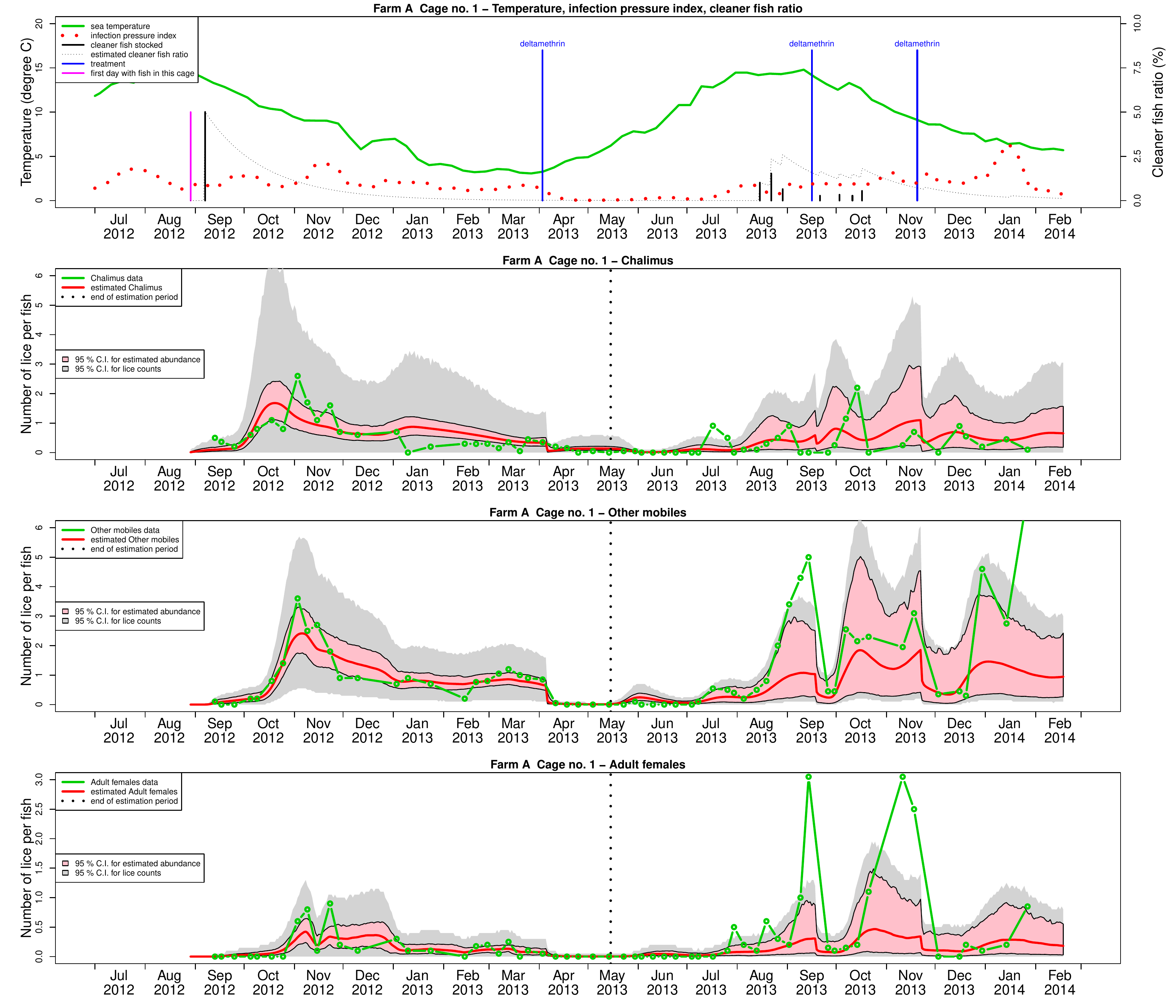}
  \caption{Fitted (until 15. May 2013) and predicted (from 16. May
    2013) values for the lice and the cleaner fish
    populations. Symbols are as given in Figure
    \ref{fig:FarmAdataonly} with the following additions: Upper panel:
    Fitted (posterior mean) cleaner fish ratio (dotted black
    curve). Three lower panels: i) Fitted values (red curves to the
    right of the vertical black dotted line), ii) predictions
    conditioned on known temperature, external infection pressure
    index and number and weight of salmon (red curves to the left of
    the vertical black dotted line), iii) corresponding 95\% credible
    interval for the lice population (pink area) and iv) additional
    95\% credible interval for lice counts (grey area), \emph{i.e.}
    including the randomness in the negative binomial distribution for
    lice counts. }
\label{fig:FarmAFittedPredicted} 
\end{center}
\end{figure}

\begin{figure}[htpb]
 \begin{center} \includegraphics[width=1\textwidth,angle=0]{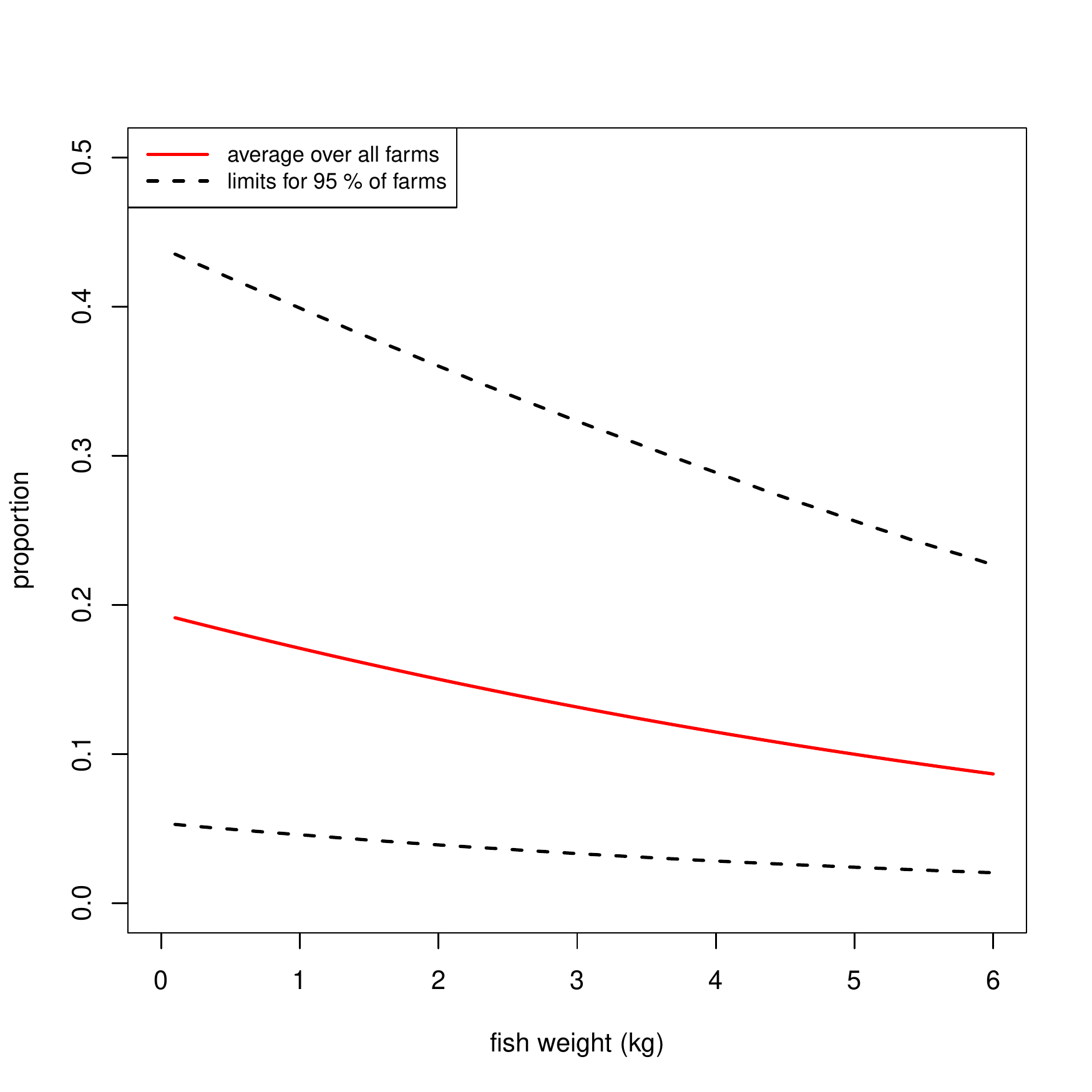}
 \caption{Posterior
 mean of the proportion of true CH lice counted as a
 function of fish weight, on average over all farms (solid line) and
 95\% limits for the between-farm variability (dashed
 lines).}  \label{fig:pCHcount}
\end{center}
\end{figure}

There was substantial underreporting of the number of lice at the CH
stage. Depending on the size of the fish, the model estimates
suggested that on average only 9\% to 19\% of the CH lice were counted
(Figure \ref{fig:pCHcount}). In addition, there was substantial
between farm variability in this counting error (Figure
\ref{fig:pCHcount}). The relationship between observed abundances of
lice at the CH stage and predicted values was poorer than for the
other stages (OM and AF), even when the underreporting was accounting
for (the grey ``counting error'' area for CH in Figure
\ref{fig:FarmAFittedPredicted} is wide and includes zero). This can be
quantified by the aggregation parameter $\rho$ which was 50-75\% lower
than for OM and AF (Table \ref{tab:estmodel.static}).  This indicates
that the information content in the counts of CH stage lice is
limited.

\clearpage
\subsection{Parameter estimates}
\label{subsec:parmest}

Posterior mean estimates and credible intervals for parameters in the
model are given in Tables \ref{tab:estmodel.static} and
\ref{tab:estmodel.timevar}. Note that for those parts of the model
where we have no data, covariation between parameters in the model may
lead to potential bias, i.e. high estimates of one parameter may be
compensated for by an associated change in the value of another
parameter. An example of this is that mortality and development rates
for the R and CO stages and the reproduction rate are related, but
without relevant observations to tease them apart.  For instance, if
the reproduction rate is overestimated, this can be compensated either
by increasing the mortality in the R and CO stages (which are assumed
to be equal) or by reducing the infection rate (development from CO to
CH).

When compared to previously published estimates on stage-specific
mortality, there are some notable differences
(Table~\ref{tab:mortality}). The estimate of the mean mortality rate
at the R and CO stages ($\lambda^{RCOnat}$) was higher than the
previous estimate (Table~\ref{tab:mortality}). However, note that in
our model, this quantity also accounts for nauplii and copepodids that
drift away from the farm, in addition to the pure natural
mortality. For the CH, PA and A stages, the mortality rates vary over
time, but we calculated their overall expectations (averages in the
long run) by simulation. The overall expectation for the mortality rate
of the CH and PA stages tended to be towards the lower range of
previous estimates, while the estimate for the adult stage was within
the range of previous studies (Table~\ref{tab:mortality}).  These
estimates must be interpreted with caution since the model assumes
equal development rates between genders and, furthermore, that adult
males are pooled together with the PA stages in the observational
data.  More detailed figures on mortality, including parameter
uncertainties, are presented in Figures 9-13 in the Supplementary
material.

\begin{table}[h!]
  \caption{Posterior means with 95\% credible intervals of parameters
    in the static parts of the model.}
  \scriptsize{
 \centering
\begin{tabular}{lllllrrr}
\hline
Part of          &     &Parameter             &Parameter         &                               &Posterior&95\% C.I.&95\% C.I. \\ 
model            &Stage&interpretation        &symbol             &Section                        &mean     &lower     &upper \\ \hline
Natural mortality&R,CO     &Mortality rate             &$\lambda^{RCOnat}$  &\ref{subsubsec:surv}            &0.303&   0.293&   0.314\\
Mortality cl.fish&PA, A    &Regression coeff.          &$\lambda^{clf}$    &\ref{subsubsec:surv}            &0.839&   0.622&   1.059\\
Development      &Egg      &Median at 10$^{\circ}$C     &$\delta^{Em10}$    &\ref{subsubsec:development}      &4.720&   4.403&   5.159\\
Development      &Nauplii  &Median at 10$^{\circ}$C     &$\delta^{Nm10}$    &\ref{subsubsec:development}      &4.096&   3.646&   4.421\\
Development      &R        &Shape parameter            &$\delta^{Rs}$      &\ref{subsubsec:development}     &18.865&  16.027&  19.975\\
Development      &R        &Power parameter            &$\delta^{Rp}$      &\ref{subsubsec:development}     &0.401&   0.400&   0.405\\
Development      &CH       &Median at 10$^{\circ}$C     &$\delta^{CHm10}$    &\ref{subsubsec:development}      &18.945&  18.329&  19.520\\
Development      &CH       &Shape parameter            &$\delta^{CHs}$      &\ref{subsubsec:development}     &8.024&   7.120&   8.976\\
Development      &CH       &Power parameter            &$\delta^{CHp}$      &\ref{subsubsec:development}     &1.299&   1.252&   1.344\\
Development      &PA       &Median at 10$^{\circ}$C     &$\delta^{PAm10}$    &\ref{subsubsec:development}      &10.700&  10.126&  11.257\\
Development      &PA       &Shape parameter            &$\delta^{PAs}$      &\ref{subsubsec:development}     &1.629&   1.445&   1.836\\
Development      &PA       &Power parameter            &$\delta^{PAp}$      &\ref{subsubsec:development}     &0.859&   0.7806&   0.937\\
Development      &CO       &Expectation                &$\delta_{0}^{CO}$   &\ref{subsubsec:developmentCO}   &-2.564&  -2.970&  -2.198\\
Development      &CO       &Regression coeff.          &$\delta^{CO}_1$     &\ref{subsubsec:developmentCO}   &0.084&   0.043&   0.125\\
Development      &CO       &Variance within farm       &$(\sigma^{COdf})^2$ &\ref{subsubsec:developmentCO}   &0.034&   0.027&   0.044\\
Development      &CO       &Variance between farms     &$(\sigma^{COd})^2$  &\ref{subsubsec:developmentCO}   &0.360&   0.202&   0.620\\
Reproduction     &AF to R  &Basic number of eggs       &$\beta^r_0$        &\ref{subsubsec:reprod}          &172.500&fixed&\\
Reproduction     &AF to R  &Age dependence             &$\beta^r_1$        &\ref{subsubsec:reprod}          &0.200  &fixed&\\
Reproduction     &AF to R  &Density dependence         &$\gamma^r$         &\ref{subsubsec:reprod}          &493    &481  & 498\\
Cleaner fish model&         &Mortality rate            &$\kappa^{clf}$      &\ref{subsubsec:cleaner.fish}    &0.027&   0.022&   0.034\\
Data model        &CH       &Aggregation parameter     &$\rho^{CH}$         &\ref{subsubsec:data.model}      &0.051&   0.049&   0.054\\
Data model        &OM=PA+AM &Aggregation parameter     &$\rho^{OM}$         &\ref{subsubsec:data.model}      &0.194&   0.183&   0.206\\
Data model        &AF       &Aggregation parameter     &$\rho^{AF}$         &\ref{subsubsec:data.model}      &0.119&   0.110&   0.131\\
Data model        &CH       &Expectation               &$\beta_{0}^{CHcount}$ &\ref{subsubsec:data.model}      &-1.572&  -1.832&  -1.345\\
Data model        &CH       &Variance                  &$(\sigma^{CHcount})^2$&\ref{subsubsec:data.model}      &0.431&   0.243&   0.730\\
Data model        &CH       &Regression coeff.         &$\beta^{CHcount}_1$   &\ref{subsubsec:data.model}      &-0.164&  -0.189&  -0.135\\
\hline
\end{tabular}
}
\label{tab:estmodel.static}\\
\end{table}

\begin{table}[h!]
  \caption{Posterior means with 95\% credible intervals of parameters in the time-varying parts of the model.}
  \scriptsize{ \centering
\begin{tabular}{lllllrrr}
\hline
Part of           &         &Parameter                &Parameter          &                               &Posterior&95\% C.I.&95\% C.I. \\ 
model             &Stage    &interpretation           &symbol             &Section                        &mean     &lower     &upper \\ \hline
Natural mortality &CH       &Expectation in AR(1)     &$\lambda_{0}^{CHnat}$&\ref{subsubsec:surv}      &-6.943&  -7.043& -6.847\\
Natural mortality &CH       &Coefficient in AR(1)     &$\phi^{CHnat}$      &\ref{subsubsec:surv}      &0.011&   0.001&  0.027\\
Natural mortality &CH       &Variance in AR(1)        &$(\sigma^{CHnat})^2$&\ref{subsubsec:surv}       &0.019&   0.011&  0.026\\
Natural mortality &PA       &Expectation in AR(1)     &$\lambda_{0}^{PAnat}$&\ref{subsubsec:surv}      &-4.908&  -5.040& -4.713\\
Natural mortality &PA       &Coefficient in AR(1)     &$\phi^{PAnat}$      &\ref{subsubsec:surv}      &0.025&   0.001&  0.065\\
Natural mortality &PA       &Variance in AR(1)        &$(\sigma^{PAnat})^2$&\ref{subsubsec:surv}       &0.130&   0.099&  0.173\\
Natural mortality &A        &Expectation in AR(1)     &$\lambda_{0}^{Anat}$&\ref{subsubsec:surv}       &-2.411&  -2.492& -2.332\\
Natural mortality &A        &Coefficient in AR(1)     &$\phi^{Anat}$      &\ref{subsubsec:surv}        &0.693&   0.676&  0.708\\
Natural mortality &A        &Variance in AR(1)        &$(\sigma^{Anat})^2$ &\ref{subsubsec:surv}       &0.729&   0.685&  0.775\\
Mortality ch.tr.  &CH, PA, A&Expectation, deltamethrin &$\lambda^{DMcht}$   &\ref{subsubsec:surv}    &2.400&   1.641&  3.145\\
Mortality ch.tr.  &CH, PA, A&Variance, deltamethrin    &$(\sigma^{DMcht})^2$&\ref{subsubsec:surv}    &9.070&   6.031& 12.427\\
Mortality ch.tr.  &PA, A    &Expectation, azamethiphos &$\lambda^{AZcht}$  &\ref{subsubsec:surv}    &0.133&  -0.675&  1.007\\
Mortality ch.tr.  &PA, A    &Variance, azamethiphos    &$(\sigma^{AZcht})^2$&\ref{subsubsec:surv}   &$(\sigma^{DMcht})^2$&&\\
Mortality ch.tr.  &PA, A    &Expectation, H$_2$O$_2$   &$\lambda^{HPcht}$   &\ref{subsubsec:surv}   &4.056&   3.219&  5.159\\
Mortality ch.tr.  &PA, A    &Variance, H$_2$O$_2$      &$(\sigma^{HPcht})^2$&\ref{subsubsec:surv}   &$(\sigma^{DMcht})^2$&&\\
Mortality ch.tr.  &CH, PA, A&Expectation, emamectin    &$\lambda^{EMcht}$   &\ref{subsubsec:surv}   &-4.744&  -5.456& -4.280\\
Mortality ch.tr.  &CH, PA, A&Variance, emamectin       &$(\sigma^{EMcht})^2$&\ref{subsubsec:surv}   &1.825&   0.865&  3.483\\
Mortality ch.tr.  &CH, PA   &Expectation, diflubenzuron&$\lambda^{DIcht}$   &\ref{subsubsec:surv}   &-8.712& -12.555& -4.638\\
Mortality ch.tr.  &CH, PA   &Variance, diflubenzuron   &$(\sigma^{DIcht})^2$&\ref{subsubsec:surv}   &$(\sigma^{EMcht})^2$&&\\
External recr.    &AF to R  &Expectation in AR(1)      &$\mu^{Ext}$        &\ref{subsubsec:omega.er}   &0.300&   0.182&  0.411\\
External recr.    &AF to R  &Coefficient in AR(1)      &$\phi^{Ext}$       &\ref{subsubsec:omega.er}   &0.934&   0.924&  0.944\\
External recr.    &AF to R  &Variance in AR(1)         &$(\sigma^{Extar})^2$&\ref{subsubsec:omega.er}   &0.164&   0.152&  0.175\\
External recr.    &AF to R  &Variance                  &$(\sigma^{Ext})^2$ &\ref{subsubsec:omega.er}   &0.007&   0.001&  0.027\\
\hline
\end{tabular}
}
\label{tab:estmodel.timevar}\\
\end{table}

\begin{table}[h!]
  \caption{Posterior means with 95\% credible intervals of daily mortality from this study together with point estimates or ranges from previous studies.}
  \scriptsize{ \centering
\begin{tabular}{lrrlll}
\hline
      &Point        &         &       &                               &        \\
      &estimate     &         &       &                               &        \\
Stage &or range     &95\% C.I.&Sex  &Comment                        &Reference\\ \hline
nauplii&0.30        &0.29-0.31&       &for R=eggs+nauplii, includes drifting away&this paper\\
      &0.17         &         &       &``plausible value''   &\cite{StienEtal2005}\\ \noalign{\vskip 1mm}
CO    &0.30         &0.29-0.31&       &same as for R                  &this paper\\
      &0.22         &         &       &''plausible values''           &\cite{StienEtal2005}\\ \noalign{\vskip 1mm}
CH    &0.0010       &0.0009-0.0011&   &                               &this paper\\
      &0.002-0.01   &         &       &''plausible values''           &\cite{StienEtal2005}\\
      &0.0006-0.020 &         &       &outer interval limits from 4 reported studies& --''--\\
      &0.0002-0.026 &         &       &range over 7 trials on juvenile Pacific salmon&\cite{KrkosekEtal2009}\\ \noalign{\vskip 1mm}
PA    &0.0079       &0.0068-0.0096&   &                               &this paper\\
      &0.02-0.18    &         &males  &''plausible values''           &\cite{StienEtal2005}\\
      &0.002-0.21   &         &males  &outer interval limits from 4 reported studies& --''--\\
      &0.03-0.07    &         &females&''plausible values''           & --''--\\
      &0.011-0.102  &         &females&outer interval limits from 4 reported studies& --''--\\
      &0.14-0.34    &         &       &PA+A combined,range over 7 trials&\cite{KrkosekEtal2009}\\
      &             &         &       &on juvenile Pacific salmon     &        \\ \noalign{\vskip 1mm}
A     &0.12         &0.11-0.13&       &                               &this paper\\
      &0.03-0.06    &         &males  &''plausible values''           &\cite{StienEtal2005}\\
      &0.008-0.26   &         &males  &outer interval limits from 4 reported studies& --''--\\
      &0.02-0.04    &         &females&''plausible values''           & --''--\\
      &0.003-0.70   &         &females&outer interval limits from 3 reported studies& --''--\\
      &0.14-0.34    &         &       &PA+A combined,range over 7 trials&\cite{KrkosekEtal2009}\\
      &             &         &       &on juvenile Pacific salmon     &\\
\hline
\end{tabular}
}
\label{tab:mortality}\\
\end{table}

\begin{figure}[htpb] \begin{center}
 \includegraphics[width=1\textwidth,angle=0]{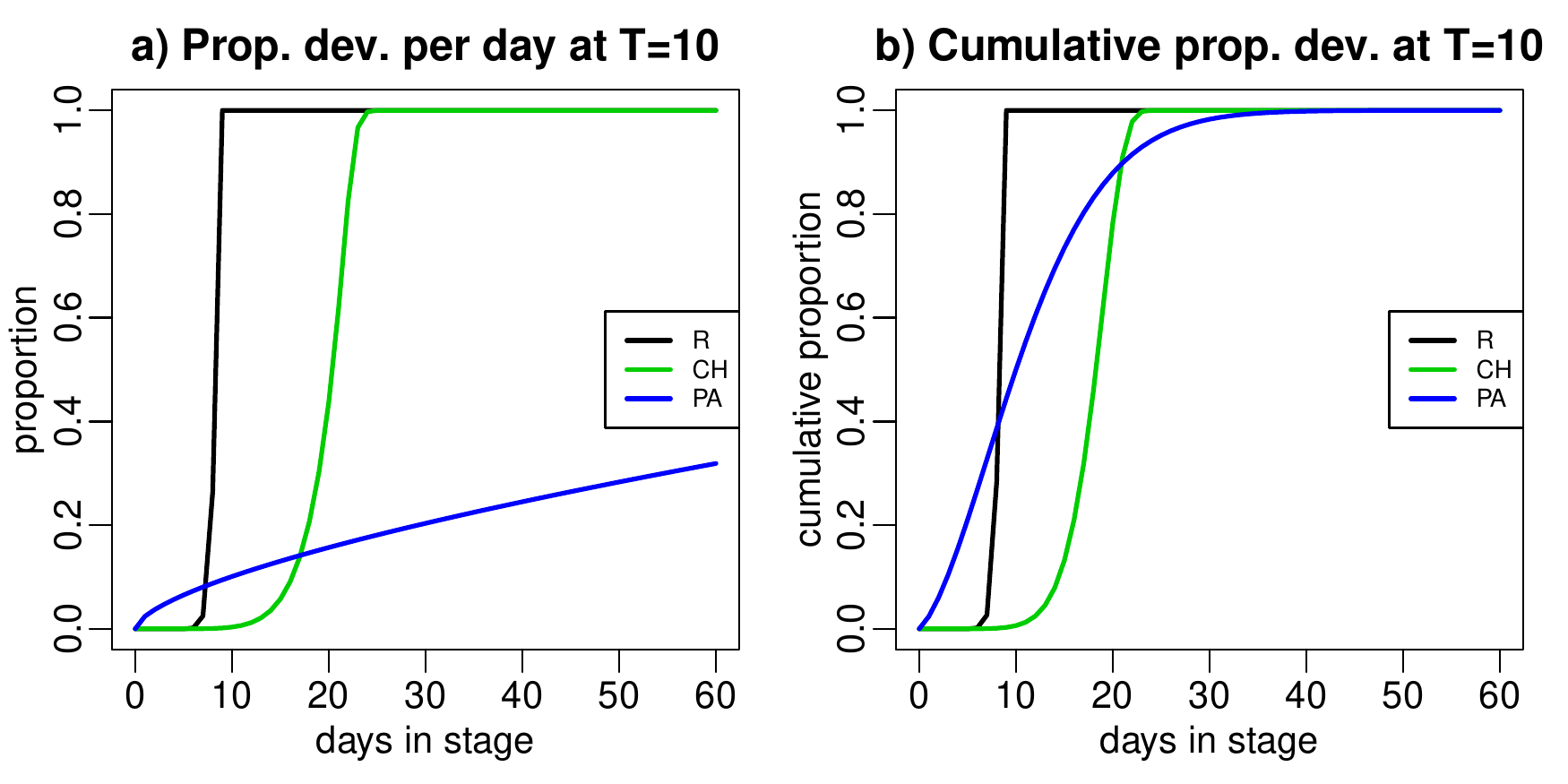}
 \caption{Posterior means of daily development (a) and cumulative
 proportion developed to next stage (b) at temperature 10 for stages
 R, CH and PA. The curves are the posterior means of the
 development rates, conditioned on each stage-age, and may therefore
 be slightly different from curves based the posterior means of the
 parameter values plugged into Eq. \eqref{eq:dev.rate}.}
 \label{fig:compactDevelopmentTimesT10} 
\end{center} 
\end{figure}

We also used simulations to find the estimated median development
times, since the $\delta^{m10}$ parameters given in Table
\ref{tab:estmodel.static} have exact interpretations only in the
continuous Weibull distribution.  Estimated median development times
at 10$^{\circ}$C were similar to mean and minimum estimates from
previous studies (Table~\ref{tab:development}), however with a
slightly higher estimate at the CH stage and fairly low estimate for
the PA stage.  Panel a) in Figure \ref{fig:compactDevelopmentTimesT10}
shows how the estimated development rates increases by stage-age at a
temperature of 10$^{\circ}$C. These curves differ in principle from
the development rates used in all population models mentioned in
Section \ref{subseq:mod.background}, since all these models use step
functions with a development rate of 0 until a minimum development
time and then a constant rate afterwards. For the R and CH stages,
however, the estimated \emph{cumulative} proportion of lice developed
to the next stage (panel b) in Figure
\ref{fig:compactDevelopmentTimesT10}) resembel these step
functions. For the PA stage, however, the estimated development rate is
fundamentally different from those used in other models, since the
estimated development rate for PA is non-zero already after one
day. This implies that some lice in the PA stage may develop to the A
stage very quickly in the present model. 

Note that these curves ignore mortality, and the cumulative
mortalities may be high for stage-ages where the development rates
still are quite low, especially at low temperatures. More detailed
figures on development times, including parameter uncertainties and
for different temperatures (5, 10 and 15$^{circ}$C), are presented in
Figures 14-16 in the Supplementary material.

\begin{table}[h!]
  \caption{Posterior means with 95\% credible intervals of development times (in days) at 10$^{\circ}$C from this study together with point estimates or ranges from previous studies.}
  \scriptsize{ \centering
\begin{tabular}{lrrllll}
\hline
       &Point        &         &        &       &                               &        \\
       &estimate     &         &Estimate&       &                               &        \\
Stage  &or range     &95\% C.I.&of what &Sex    &Comment                        &Reference\\ \hline
eggs   &4.8          &4.5-5.3  &median  &       &                               &this paper\\
       &8.8          &         &minimum &       &from their Eq. (8) and Table 3 &\cite{StienEtal2005}\\
       &4.6          &         &mean    &       &                               &\cite{SamsingEtal2016}\\ \noalign{\vskip 1mm}
nauplii&4.2          &3.7-4.5  &median  &       &                               &this paper\\
       &3.6          &         &minimum &       &from their Eq. (8) and Table 3 &\cite{StienEtal2005}\\
       &3.8          &         &mean    &       &                               &\cite{SamsingEtal2016}\\ \noalign{\vskip 1mm}
CH     &18.8         &18.0-19.0&median  &       &                              &this paper\\
       &15.4         &         &minimum &males  &from their Eq. (8) and Table 3&\cite{StienEtal2005}\\
       &16.5         &         &minimum &females&from their Eq. (8) and Table 3&\cite{StienEtal2005}\\
       &11-13        &         &range   &males  &                              &\cite{EichnerEtal2015}\\
       &13-15        &         &range   &females&                              &\cite{EichnerEtal2015}\\
       &11-14        &         &minimum &       &5 trials on juvenile Pacific salmon at 9-11$^\circ$C&\cite{KrkosekEtal2009}\\ \noalign{\vskip 1mm}
PA     &10.5         &10.0-11.0&median  &       &                              &this paper\\
       &10.4         &         &minimum &males  &calculated as difference of their time&\cite{StienEtal2005}\\ 
       &             &         &        &       &from CH to A and from CH to PA  &\\
       &15.4         &         &minimum &females&calculated as difference of their time&\cite{StienEtal2005}\\ 
       &             &         &        &       &from CH to A and from CH to PA  &\\
\hline
\end{tabular}
}
\label{tab:development}\\
\end{table}

\begin{figure}[htpb]
 \begin{center} \includegraphics[width=1\textwidth,angle=0]{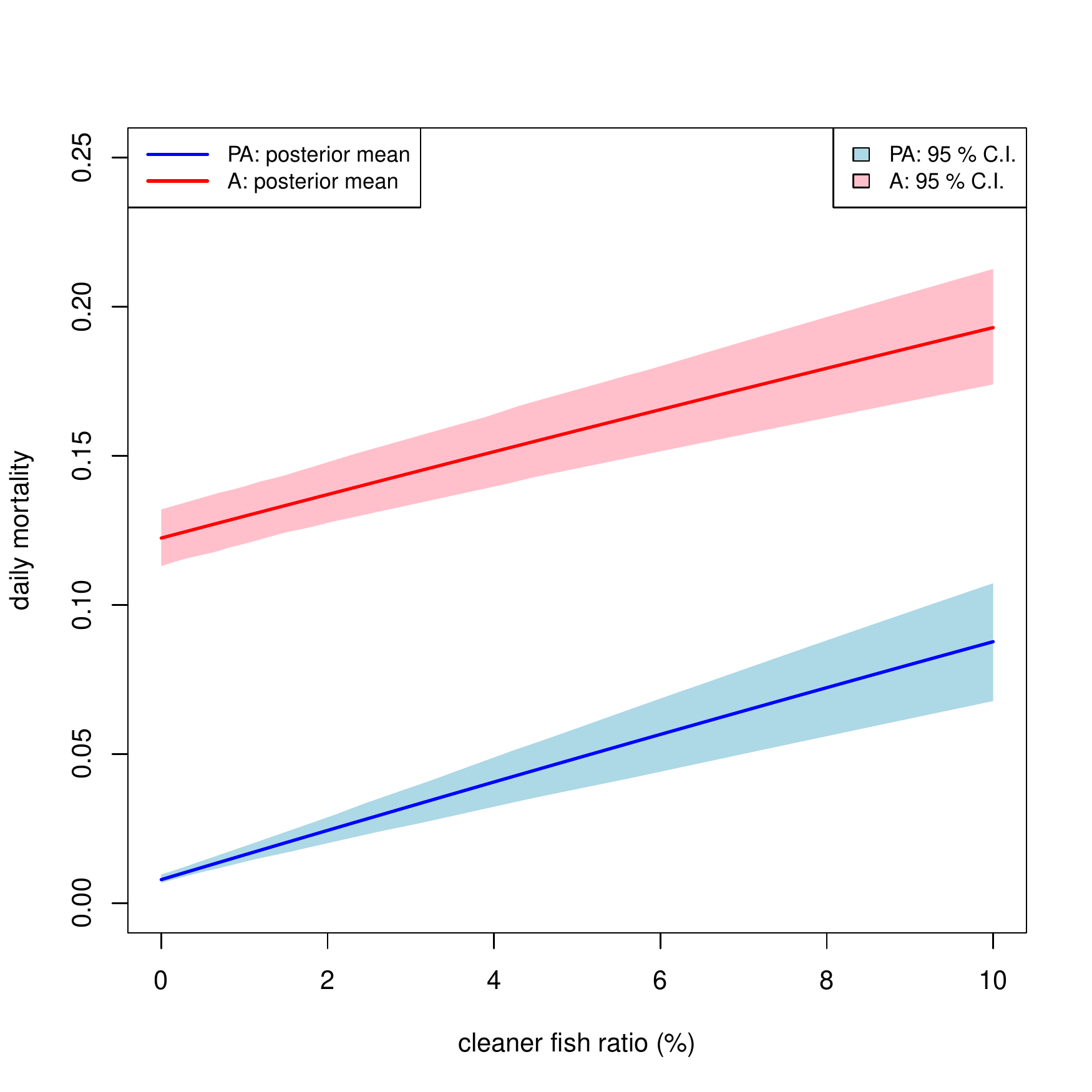}
 \caption{Posterior
 mean and 95\% credible intervals for the total mortality for the PA
 and A stages when the cleaner fish ratio is increased from 0 (only
 natural mortality) to 10\%.}
 \label{fig:cleanerfishmort}
\end{center}
\end{figure}

We obtain an estimate of the daily cleaner fish mortality of 0.027
(C.I. 0.022-0.034, Table \ref{tab:estmodel.static}). This suggest that
the cleaner fish population is reduced to its half about 1 month after
release. The model confirms that there is increased mortality of lice
associated with the use of cleaner fish
(Figure~\ref{fig:cleanerfishmort} and
Table~\ref{tab:estmodel.static}). With a 10\% cleaner fish to salmon
ratio, the estimated daily lice mortality (for lice in the PA and A
stages) due to the use of cleaner fish, is 0.080
(C.I. 0.060-0.100). This implies a reduction in the life expectancy for
adult lice from 8.5 to 5 days with an increase in cleaner fish ratio
from 0 to 10\%, and a decrease in the life expectancy for pre-adult
lice (PA) going from 127 days to 11 days for the same change in
cleaner fish ratio. Therefore, in particular for PA lice, the use of
cleaner fish is estimated to have a substantial effect on lice
survival. However, in the present data, the (estimated) cleaner fish
ratio seldom amounted stocked to more than 5\%, which seems to be too low
to avoid additional treatments, since medical treatments were applied
in almost all cages in the data set.

The effects of the various chemotherapeutic treatments are difficult
to compare since the assumed duration of the effects varies between
treatments and by temperature, and because they affect different
stages of lice. Furthermore, since lice develop resistance
towards such treatments \citep{AaenEtal2015}, we expect that the
effect will decrease over time. Nevertheless, the estimated expected
cumulative mortality of lice (found by simulation) in the PA or A
stages due to bath treatments (\emph{i.e.} non-feed) ten days post
treatment at 10$^{\circ}$C, were high for hydrogen peroxide (0.99,
C.I. 0.97-1.00), and deltamethrin/cypermethrin (0.94 , C.I. 0.88-0.97)
and somewhat lower for azamethiphos (0.74 , C.I. 0.64-0.85). The first
two of these are similar to what others have reported for
non-resistant lice populations, being 99\% for hydrogen peroxide
\citep{GronerEtal2013} and 95\% for deltamethrin and cypermethrin
\citep{RevieEtal2005}.  The estimated effects of all treatment types
are further illustrated in Figures 9-13 in the Supplementary material.

\begin{figure}[htpb]
 \begin{center} \includegraphics[width=1\textwidth,angle=0]{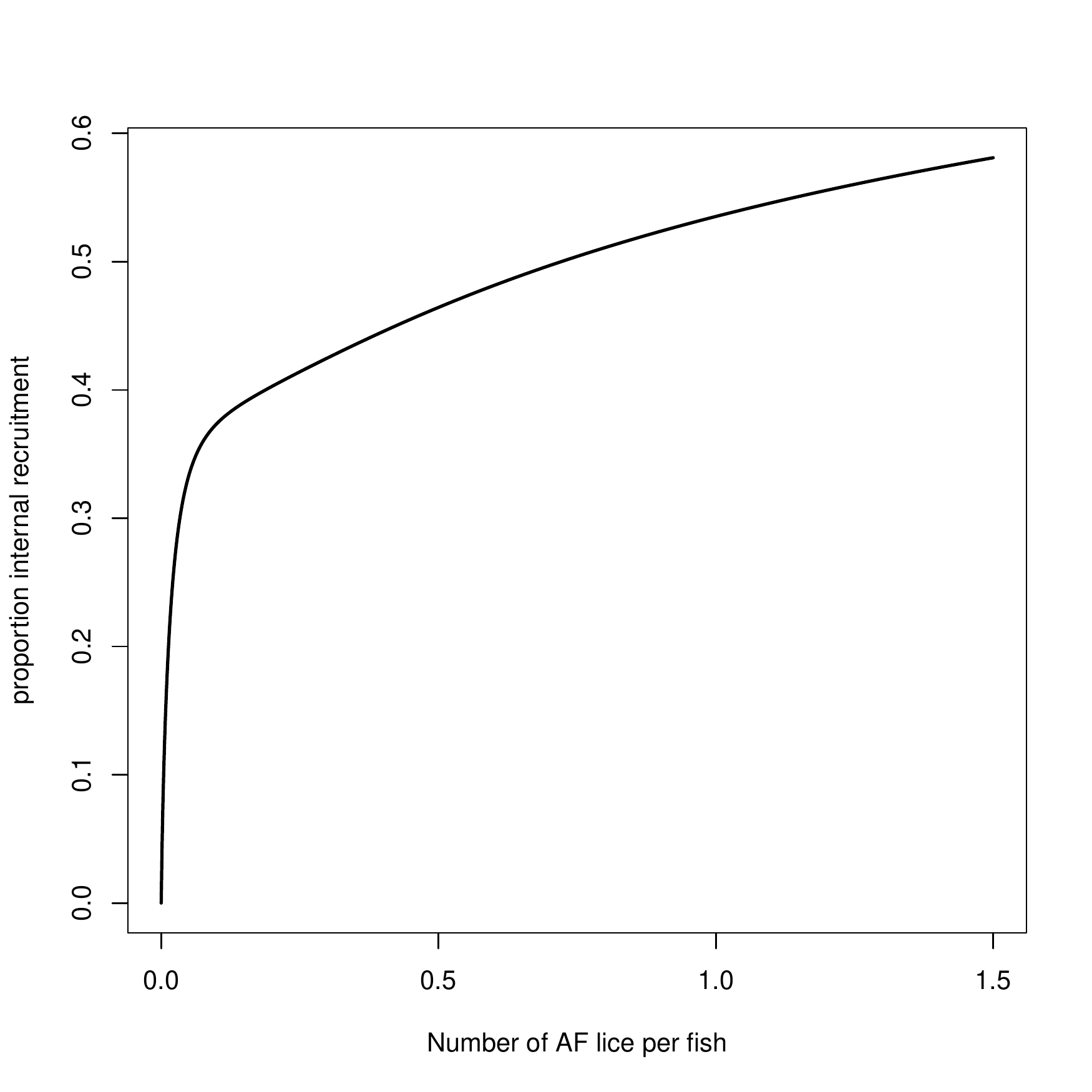}
   \caption{Posterior mean of the average proportion of recruitment
     that has an internal source as a function of abundance of adult
     females, averaged over all farms and all time points.}
\label{fig:intProportion} 
\end{center}
\end{figure}

Both external and internal recruitment varied substantially over
time, and one of the two may dominate the other in certain
periods. The estimated proportion of internal recruitment for each
farm averaged over the whole production cycle varies from about 4 to
73\%. On average over all farms, this proportion is 24\%
(C.I. 23-26\%), with a median of 19. This is in accordance with
results in \cite{Aadlandsvik2015}, who reported a median of 18\% for
the proportion of internal recruitment, based on a simulation of the
spread of lice larvae between 591 farms from a hydrodynamic model that
takes sea currents into account \citep{JohnsenEtal2014}. When the
abundance of adult female lice was low in the farm, naturally internal
recruitment tended to be low, while internal recruitment was estimated
to be very important in farms with a high abundance of adult females
(Figure~\ref{fig:intProportion}).

Concerning the reproduction rate, we notice that the estimate of
$\gamma^r_0$ is 493 (C.I 481-498). In practice, this means that there
is no evidence of density dependent reproductive rates (Allee effect)
in the estimated model, contrary to the hypothesis suggested by
\cite{StormoenEtal2013}, \cite{KrkosekEtal2012} and
\cite{GronerEtal2014}. One reason for this discrepancy is that
recruitment from external sources is not considered in the
aforementioned models. In our model, internal recruitment is on
average of less importance than external recruitment at low abundances
of AF lice. Reduced reproductive rates at low abundances of infection
are therefore likely to be masked by external recruitment.

\clearpage
\section{Conclusion}
\label{sec:conclusion}

The presented process model for the population dynamics of salmon lice
in aquaculture farm systems combine 1) a model for the main stage
structure of salmon lice; 2) model terms that describe internal and
external recruitment of salmon lice to the farms; 3) models for the
impact of a management strategies adopted to reduce and control salmon
lice infection levels; and 4) allows for stochasticity in aspects of
these processes. In addition, we describe a model for the link between
the process model and data collected on a routine basis in modern
aquaculture. This allows the process model to be fitted to data and
updated when new production and salmon lice data from fish farms
become available. The model produces estimates of lice abundances in
fish farms that correspond well with observed infection
levels. Furthermore, it allows reasonable predictions of pre-adult and
adult sea lice infection levels to be made for several weeks into the
future (Figure~\ref{fig:FarmAFittedPredicted}). We note, however, that
the observational data on the number of lice at the chalimus stage
vary substantially, and is underestimated to varying degrees between
fish farms. This suggests that the data collected on the abundance of
lice at the chalimus stages contain limited information in terms of
real infection levels in the farms.

The model generates stage specific estimates of development and
mortality rates based on real production data. For the planktonic
stages, the estimates of mortality rates were high relative to earlier
field and laboratory studies, but these are not directly comparable,
since our definition of mortality include lice that drift away from
the farm.  For the chalimus and pre-adult stages, the mortality rates
were low compared to previous studies, while it was rather high for
the adult stage. One reason could be the poor data support for the
early life stages, since we have no data on the planktonic stages and
poor data on the chalimus stages. The poor support by data makes it
difficult to separate survival-estimates at these stages from other
parameters in the model, in particular reproduction rates, development
rates and biases in chalimus counts, and perhaps this problem
propagates to the pre-adult stage.  The median development times for
are reasonable. However, we note that the distribution of development
times from the pre-adult to the adult stages include the possibility
of very short development times.  It is unclear why this happens, but
one possible explanation could be movement of adult lice between cages
within a fish farm. Another explanation could be that some of the
non-gravid female adults have been misclassified as pre-adults in the
lice count data.

The model is well suited to quantify the effect of different
treatments on lice infection levels. This aspect may be useful in
monitoring development of resistance to treatments in sea lie
populations. Furthermore, to our knowledge this is a first data-based
quantification of the effect of using cleaner fish to manage lice
levels in full scale salmon farms.

The model allows both internal and external infection processes to be
quantified. When there are no adult female lice in a fish farm, all
new infections will necessarily come from external sources. When
reproducing adult female lice are present, however, they may
contribute substantially to infection pressure at the farm. Their
impact on local recruitment will depend both on their numbers and
their reproductive rate, but also on the external infection
pressure. The model did not suggest any evidence of density dependent
reproductive rates.


The emphasis on salmon louse control in salmon farming has gravely
intensified in later years, resulting in the implementation of a
variety of innovative control methods. Some of these methods aim to
reduce salmon louse recruitment from external sources, whereas other
methods aim at increasing the mortality of parasitic stages of the
lice, such as the use of cleaner fish. To tease out actual effects of
such control efforts at farm levels, or more so for farms in a
production area, is not a simple task, given the interconnected
production-system of fish farms and planktonic spread of this
parasite.  The present model for the population dynamics of salmon
lice in aquaculture farm systems may resolve the complexity of
processes to a degree that will improve evaluations of effects of
various control efforts at farm levels. We believe that using the
model as ``mathematical laboratory'' to explore effects of different
salmon louse control strategies in interconnected farms in production
areas has a large potential for rationalising area-wise control
strategies in an informed manner.

\section* {Acknowledgements}

This work was funded by the Norwegian Seafood Research Fund through
the project FHF 900970 ``Populasjonsmodell for lakselus p{\aa} merd og
lokalitetsniv{\aa}''.  We thank the fish farming companies Marine Harvest,
Salmar and M{\aa}s{\o}val for supplying us with their detailed
production and lice count data.

\clearpage

\bibliography{sealice}

\bibliographystyle{environmetrics}




\end{document}